%% file: main.tex
\documentclass[sigconf]{acmart}
\renewcommand\footnotetextcopyrightpermission[1]{}
\usepackage{bm}
\usepackage{enumitem}
\usepackage{booktabs}
\usepackage{adjustbox}
\usepackage{multirow}
\usepackage{svg}
\usepackage{graphicx} 
\usepackage{makecell}
\usepackage{algorithmic}
\usepackage{ulem}
\usepackage{pifont}
\usepackage{xcolor}
\usepackage{subcaption}
\usepackage[dvipsnames]{xcolor}

\AtBeginDocument{%
  }

\setcopyright{none}
\newcommand{\vpara}[1]{\vspace{0.05in}\noindent \textbf{#1 }}

\newcommand{\lyy}[1]{#1}

\begin{document}

\title{IDSTune: A Multi-Agent Collaborative Framework for Integrated Database System Tuning\\}


\author{Yiyan Li\textsuperscript{*}, Guanli Liu\textsuperscript{\textdagger}, Renata Borovica-Gajic\textsuperscript{\textdagger}, Haoyang Li\textsuperscript{*}, Zihang Qiu\textsuperscript{*}, Xinmei Huang\textsuperscript{*},\\ Andreas Kipf\textsuperscript{\textdaggerdbl}, Cuiping Li\textsuperscript{*}, Hong Chen\textsuperscript{*}}
\affiliation{%
  \institution{\textsuperscript{*} Renmin University of China, \textsuperscript{\textdagger} The University of Melbourne, \textsuperscript{\textdaggerdbl} University of Technology Nuremberg}
  \country{}
}
\email{{liyiyan, lihaoyang.cs, qiuzihang2024, huangxinmei, licuiping, chong}@ruc.edu.cn}
\email{{guanli.liu1, renata.borovica}@unimelb.edu.au}  \email{andreas.kipf@utn.de}









\renewcommand{\shortauthors}{Li et al.}

\begin{abstract}
Database tuning is critical for achieving high performance in modern database management systems (DBMSs). 
Existing methods typically optimize a single component---knobs, indexes, or materialized views---without accounting for their interdependencies. 
\lyy{This limitation arises because these components require different tuning strategies and are difficult to integrate within a unified framework. As a result, directly extending a method to multiple components or simply combining separate methods often fails to capture cross-component collaboration and shared tuning signals. Moreover, existing methods are insufficient for handling diverse workloads, evolving data, and dynamic query patterns.} 

To address these limitations, we propose \textbf{IDSTune}, an integrated tuning framework that jointly optimizes multiple configuration components through LLM-driven multi-agent collaboration. 
IDSTune operates in two phases: \textit{(i) workload compression}, which extracts and selects task-relevant features, and \textit{(ii) configuration recommendation}, where specialized agents collaboratively generate and refine configurations for knobs, indexes, and materialized views under the supervision of a centralized coordinator. 
By incorporating feedback and external knowledge retrieval, IDSTune achieves efficient and globally consistent tuning.
\lyy{Extensive experiments show that IDSTune achieves up to 38\% performance improvement and 57\% faster tuning, with  strong adaptability across diverse scenarios.}
\end{abstract}



\vspace{-2em}
\keywords{Database Tuning, Large Language Models, Physical  Design, Knobs}
\vspace{-2em}



\maketitle

\input{intro}
\input{background}

\input{overview}
\input{workload_compression}
\input{configuration_recommendation}

\input{evaluation}

\vspace{-0.5em}
\section{Conclusion}
\label{sec:conclusion}
This paper presents IDSTune, an LLM-driven multi-agent collaborative framework for integrated database system tuning, representing the first solution to jointly optimize knobs, indexes, and MVs in a unified framework. In contrast to traditional methods that tune these components in isolation, IDSTune provides a more adaptable solution for heterogeneous database systems, exhibits stronger tolerance to data and workload drift, and better supports both OLAP and OLTP settings.  
Evaluated against other state-of-the-art baselines under diverse scenarios, IDSTune discovers configurations that improve performance by up to 38\% over the next best approach.

\begin{acks}
The authors used ChatGPT to assist with English editing and code development. No experimental results or original contributions were generated solely by AI tools.
\end{acks}

\balance
\bibliographystyle{ACM-Reference-Format}
\bibliography{myref}

\clearpage
\appendix
\input{appendix}

\end{document}

%% file: intro.tex
\section{Introduction}
\label{sec:intro}

\lyy{Modern database management systems (DBMSs) expose numerous tunable components, typically categorized into three types: knobs, indexes, and materialized views (MVs)~\cite{Chaudhuri2007@selftuningsystem}.}
Knobs typically correspond to system-level
parameters that influence database behavior, whereas
indexes and MVs represent discrete physical design decisions that directly
modify the database structure and expand the query plan search space, often incurring non-trivial creation and maintenance costs.
Together, these components define the DBMS configuration, which determines how queries are executed and how efficiently the system performs. 
Finding the optimal configuration for a given workload is challenging due to the high-dimensional nature of these components and their complex interdependencies~\cite{Zhang24@Holon}. 



\lyy{Over the past decade, extensive studies have explored single-component tuning~\cite{Li25@AgentTune, lao24@gptuner, Ji25@LIOF, siddiqui23@indextuningsurvey, Xu24@Uniview, Elena97@MVselection}, including knobs~\cite{ Li25@AgentTune, lao24@gptuner}, indexes~\cite{Ji25@LIOF, siddiqui23@indextuningsurvey}, and MVs~\cite{Xu24@Uniview, Elena97@MVselection}. While these approaches effectively optimize individual components, they remain limited in scope and fail to capture dependencies across different DBMS subsystems.}


\begin{figure}[t]
    \centering
    \includegraphics[width=0.48\textwidth]{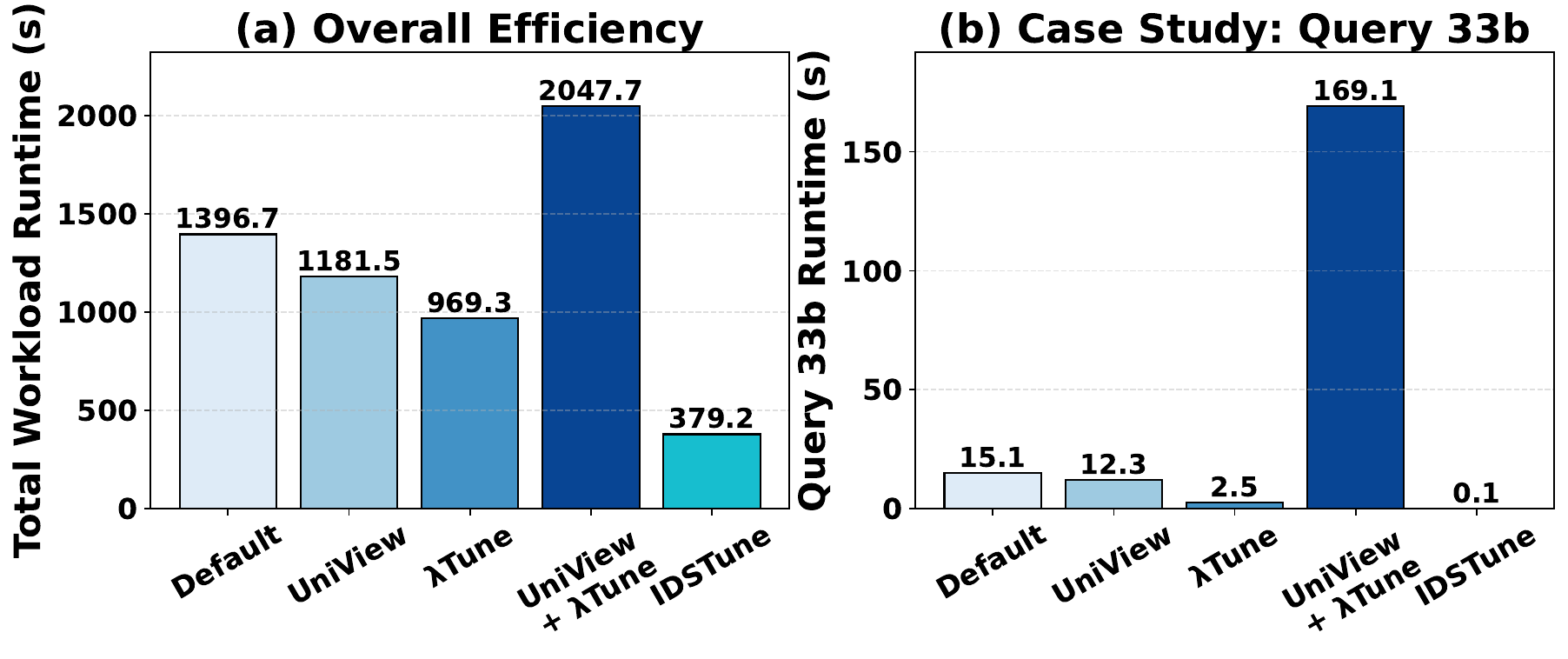}
      \vspace{-2.7em}
    \caption{\lyy{Combining UniView (materialized views recommendation) and $\lambda$Tune (knob and index recommendation) results in a slowdown, while IDSTune improves performance on the JOB benchmark: (a) overall performance, and (b) Query 33b.}}
    \label{fig:performance_comparison}
    \vspace{-2.5 em}
\end{figure}

\lyy{To address this,} recent research has extended beyond single-component tuning to jointly optimize multiple configurations~\cite{Zhang24@Holon, Immanuel2025@lambdatune, Perera22@HMAB}. 
Existing studies can be broadly categorized into two directions: 
(1) \textit{Knob+Index} tuning, which investigates the interaction between system parameters and indexes~\cite{Zhang24@Holon, Immanuel2025@lambdatune}; and 
(2) \textit{MV+Index} tuning (often referred to as physical design tuning), which explores the joint effects of indexes and materialized views~\cite{Perera22@HMAB}. 

\lyy{Although these methods achieve promising results within their respective configuration spaces, they essentially target only a \textbf{subset} of the database configuration spectrum. 
As a result, they struggle to generalize to real-world scenarios involving multiple configuration types. 
Moreover, they lack coordination mechanisms to reconcile conflicting recommendations across components, often leading to inconsistent or suboptimal configurations.
For example, in the JOB benchmark~\cite{Leis15@JOB}, combining $\lambda$-Tune~\cite{Immanuel2025@lambdatune} (a leading knob and index tuning algorithm) with UniView~\cite{Xu24@Uniview} (a state-of-the-art MV recommendation method) leads to degraded performance, as shown in Fig.~\ref{fig:performance_comparison}(a). This effect is particularly evident in query 33b (Fig.~\ref{fig:performance_comparison}(b)), where conflicting optimizations—UniView materializing an intermediate result and $\lambda$-Tune increasing \texttt{work\_mem}—cause a plan change and substantial unnecessary computation. This highlights the need for an integrated framework that  jointly optimizes multiple components and resolves cross-component conflicts.}

Extending existing tuning methods to additional components reveals three fundamental limitations:
\begin{itemize}[leftmargin=2.1em, itemsep=0.3em, topsep=0.3em]

\item[(L1)] \textit{Limited optimization effectiveness.} 
Different configuration components are highly interdependent, and optimizing them in isolation often produces conflicting adjustments. As a result, combining locally optimal solutions may fail to achieve global optimality or even degrade performance.

\item[(L2)] \textit{Inefficient tuning process.} 
Even single-component ML-based methods~\cite{Li19@qtune, Zhang21@CDBTune} often require hundreds of iterations to converge. 
Each iteration involves executing the workload (i.e., workload replay), leading to tuning processes lasting for hours. 
When multiple tuning algorithms are combined, cumulative training and evaluation costs become prohibitively high.

\item[(L3)] \textit{\lyy{Limited scenarios.}}
\lyy{Existing approaches are usually designed for fixed configuration spaces, specific database engines, and static workloads, making them difficult to apply to complex, diverse, and dynamic real-world production scenarios.} 
\end{itemize}

To address these challenges, we propose \textbf{IDSTune} (\uline{I}ntegrated \uline{D}atabase \uline{S}ystem \textbf{Tuning}), a framework driven by large language model (LLM)-based multi-agent collaboration. 
At its core, a centralized supervisor coordinates multiple specialized agents, detects conflicts, and enforces globally consistent configuration decisions.

IDSTune decomposes the tuning process into two main phases: 
\textit{(i) workload compression}, which extracts and filters task-relevant workload features; and 
\textit{(ii) configuration recommendation}, where specialized agents collaboratively optimize knobs, indexes, and MVs under centralized supervision. 
This coordinated multi-agent design enables holistic optimization across interdependent configuration components (addressing L1), while its iterative feedback mechanism accelerates convergence and improves tuning efficiency (addressing L2). 
\lyy{Additionally, rich features, training-free design, and search capability enable our method to be rapidly deployed and adapt to diverse application scenarios (addressing L3).}

Our work makes the following contributions:

\begin{itemize}
[leftmargin=1.1em, itemsep=0.3em, topsep=0.3em]
\item We propose \textbf{IDSTune}, the first LLM-driven multi-agent framework for integrated database tuning, jointly optimizing knobs, indexes, and MVs within a unified process.

\item We design a centralized multi-agent architecture with built-in search, where heterogeneous agents collaboratively refine tuning decisions through coordinated interaction and feedback, handling cross-component complexity.

\item We introduce a feature-based workload compression paradigm that adaptively selects workload- and task-relevant features, providing compact, informative tuning contexts.

\item We conduct extensive experiments on seven workloads spanning OLTP, OLAP, and real-world scenarios, demonstrating that \textbf{IDSTune} consistently outperforms state-of-the-art baselines in effectiveness, efficiency, and robustness, with detailed ablations and analysis of LLM-related overhead.
\end{itemize}

The remainder of this paper is organized as follows.
Section~\ref{sec:background} reviews related work on database system tuning, including ML- and LLM-based approaches. 
Section~\ref{sec:overview} provides an overview of the IDSTune framework. 
Section~\ref{sec:workload compression} details the workload compression phase, while Section~\ref{sec:configuration_recommendation} presents the multi-agent tuning architecture. 
Section~\ref{sec:experiments} reports experimental results and ablation studies, and Section~\ref{sec:conclusion} concludes the paper.

%% file: background.tex
\vspace{-0.5em}
\section{Background and Related Work}\label{sec:background}

Database system tuning has long been recognized as a key factor in achieving high query performance.  
A wide range of configurable components, including knobs, indexes, and MVs, collectively determine query efficiency and resource utilization.  
Early approaches relied on heuristic rules and cost models embedded in the DBMS optimizer~\cite{pgtune, Zhu17@BestConfig}, but they often failed to adapt to complex workloads and evolving environments.  
Over time, research has evolved from rule-based heuristics~\cite{mysqltuner} to learning-based~\cite{Kanellis22@llamatune} and, more recently, LLM-enhanced tuning methods~\cite{lao24@gptuner}.  
We here review representative ML- and LLM-based approaches for database tuning.


\begin{figure*}
\vspace{-1em}
    \centering
    \includegraphics[width=0.96\textwidth]{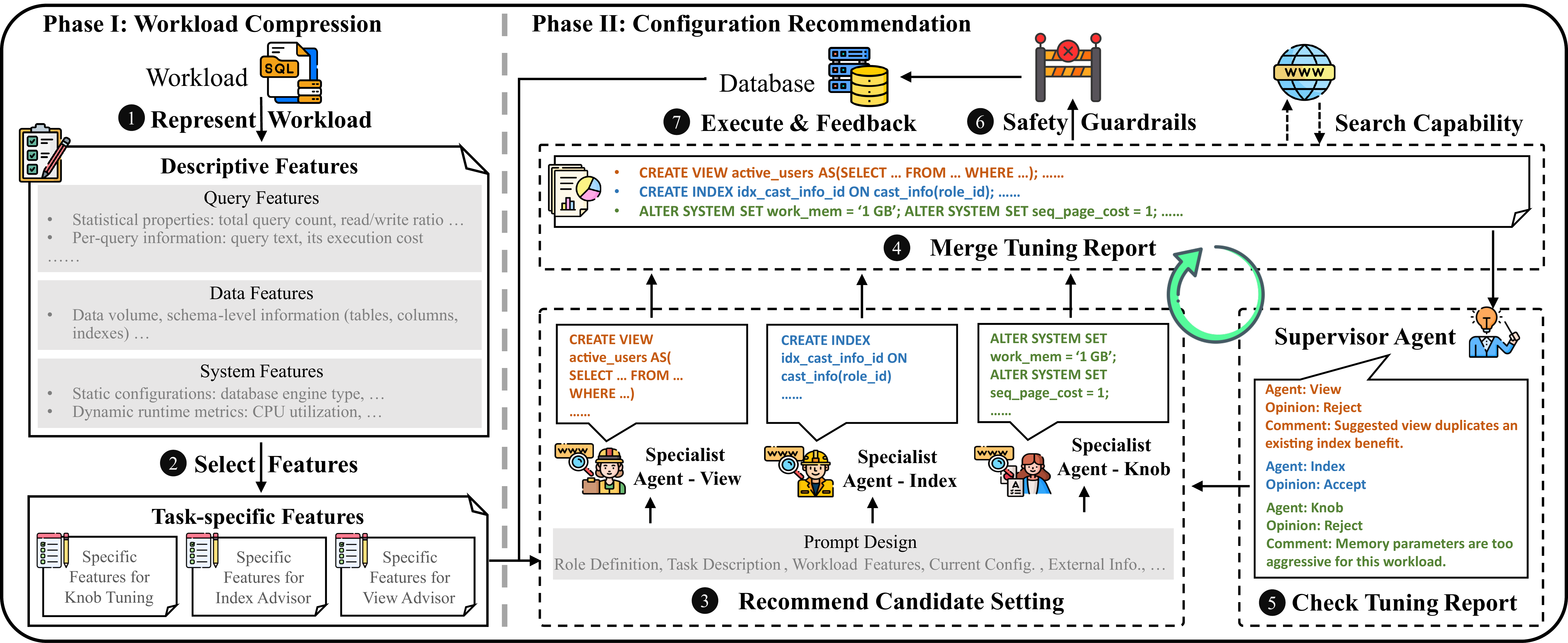}
      \vspace{-1em}
    \caption{\lyy{Overview of the two phases of IDSTune. In \textit{Phase~\uppercase\expandafter{\romannumeral 1} (Workload Compression)}, the input workload is first transformed into structured features and then filtered by a \textit{Selection Agent} for downstream tuning. In \textit{Phase~\uppercase\expandafter{\romannumeral 2} (Configuration Recommendation)}, multiple \textit{Specialist Agents} (e.g., knob, index, and view specialists) independently generate candidate configurations based on the compressed workload. Their outputs are iteratively consolidated and refined by a \textit{Supervisor Agent} to resolve conflicts and ensure completeness. A hybrid \textit{Safety Guardrails} module further validates the final configuration before deployment.}}
    \label{fig:overview}
    \vspace{-1em}
\end{figure*}

\vspace{-1em}
\subsection{ML-Based Approaches}
In recent years, a large body of research has applied machine learning techniques to database system tuning~\cite{Li19@qtune, Aken17@OtterTune, Zhang22@OnlineTune, Ji25@LIOF, Perera22@HMAB}.  
These methods learn the relationship between workloads and DBMS performance metrics and iteratively refine tuning decisions through feedback.  
Representative systems such as OtterTune~\cite{Aken17@OtterTune}, CDBTune~\cite{Zhang21@CDBTune}, and QTune~\cite{Li19@qtune} apply ML and reinforcement learning (RL) to recommend knob settings.  
Other studies target physical design tasks such as index recommendation~\cite{Ji25@LIOF} and MV selection~\cite{Xu24@Uniview}, demonstrating the effectiveness of learning-based tuning.

More recent work explores the joint optimization of multiple configuration types to capture cross-component dependencies. For example, several studies investigate \textit{knob+index} co-tuning~\cite{Zhang24@Holon, Immanuel2025@lambdatune} or \textit{MV+index} co-tuning~\cite{Perera22@HMAB}, modeling interactions between configuration elements.
While these approaches represent an important step toward holistic tuning, they are limited to pairwise combinations and specific workloads.
This motivates the need for an integrated tuning framework that can jointly consider multiple configuration spaces in a coordinated and adaptive manner.



\vspace{-0.8em}
\subsection{LLM-Based Approaches}
Large language models have demonstrated strong generalization and reasoning capabilities across a wide range of complex optimization tasks~\cite{Zhou24@dbot, devlin18@bert, huang2025@E2ETune, lao24@gptuner}.  
Unlike traditional ML models that require extensive domain-specific training, LLMs can integrate contextual knowledge and infer relationships among interdependent factors~\cite{Li24@LLM4DB}.  
These capabilities make LLMs particularly suitable for database system tuning~\cite{li2024@LLMTuningSurvey}, which demands understanding workload semantics, navigating large configuration spaces~\cite{Zhao23@tuningsurvey}, and balancing trade-offs among multiple objectives~\cite{Georgoulakis22@multiobjOptimization}.

Motivated by these strengths, recent studies have begun applying LLMs to data management tasks.  
Early efforts focus on natural language interfaces, such as text-to-SQL translation~\cite{li2024@codes, Fan24@text2sql} and query rewriting~\cite{Liu24@rewrite, Li24@LLMR2}, where LLMs demonstrate strong reasoning and contextual understanding of SQL queries.  
Beyond query-level understanding, LLMs have also been explored for system-level tasks such as performance diagnosis~\cite{Zhou24@dbot} and index recommendation~\cite{li25@indexAgent}.  

More recently, researchers have started integrating LLMs into database tuning~\cite{li2024@LLMTuningSurvey}.
For instance, AgentTune~\cite{Li25@AgentTune} and LLMIdxAdvis~\cite{zhao25@llmidxadvisre} leverage LLMs for knob tuning and index recommendation, respectively, while $\lambda$-Tune employs an LLM to jointly recommend knobs and indexes.
However, these systems rely on the most straightforward form of LLM invocation, treating the LLM as a single-call decision maker for one or two configuration types, without coordination among multiple agents.
As a result, they fail to exploit the potential of LLMs for multi-component database optimization.
Moreover, their inputs are static and predefined, lacking the ability to retrieve external domain knowledge from the web, which further limits adaptability and performance~\cite{Li25@AgentTune}.

%% file: overview.tex
\vspace{-0.5em}
\section{Overview}~\label{sec:overview}
\textbf{IDSTune} is an LLM-driven multi-agent collaborative framework for database configuration recommendation.  
As shown in Fig.~\ref{fig:overview}, the framework operates in two phases: 
\textbf{Phase~\uppercase\expandafter{\romannumeral 1}}~(Workload Compression) and 
\textbf{Phase~\uppercase\expandafter{\romannumeral 2}}~(Configuration Recommendation).  
The following paragraphs summarize the key ideas of each phase, while the detailed mechanisms are described in 
Sections~\ref{sec:workload compression} and~\ref{sec:configuration_recommendation}.

\textbf{Phase~\uppercase\expandafter{\romannumeral 1}: Workload Compression.}
IDSTune first \ding{202} \textit{represents} the input workload as a set of descriptive features capturing SQL query patterns, data distribution, and runtime statistics (Section~\ref{sec:workload representation}).  
Since different agents 
emphasize different workload 
characteristics, IDSTune adapts feature selection accordingly.  
For instance, OLAP workloads prioritize features such as join complexity and aggregation depth, whereas OLTP workloads rely more on transaction throughput and contention.  
To prevent irrelevant information from diluting the LLM’s attention,  
IDSTune \ding{203} employs a \textit{Selection Agent} to filter relevant features for each tuning task (Section~\ref{sec:workload selection}).

\textbf{Phase~\uppercase\expandafter{\romannumeral 2}: Configuration Recommendation.}
After workload compression, IDSTune performs integrated configuration optimization through a centralized multi-agent framework composed of \textit{Specialist Agents} and a \textit{Supervisor Agent} (Section~\ref{sec:configuration_recommendation}).  
\ding{204} Each specialist agent focuses on a specific configuration component (e.g., knobs, indexes, or materialized views) and independently proposes candidate settings.  
\ding{205} Their outputs are merged into a unified tuning report, which is then reviewed by the supervisor agent.  
\ding{206} The supervisor checks for potential conflicts, redundancies, and missing configurations, and requests revisions when necessary.  
This iterative feedback process continues until a coherent and optimized configuration report is produced.
\ding{207} Before deployment, the tuning report is rigorously validated by our hybrid \textit{Safety Guardrails} component.
The combination of white-box rule-based constraints and black-box LLM-based verification serves as a final safeguard to filter out invalid or risky parameters.
Any detected violation is immediately fed back to the agents for correction (Section~\ref{sec: safety guardrails}).
\ding{208} The final configuration is subsequently applied to the DBMS for evaluation, and the resulting performance feedback is incorporated into future iterations.  
Inspired by REACT~\cite{Yao2023@ReAct}, all agents in IDSTune are also equipped with web-search capabilities to retrieve external domain knowledge before taking actions (Section~\ref{sec:search_capability}).

%% file: workload_compression.tex
\section{Workload Compression}
\label{sec:workload compression}
LLM-based tuning requires a task-aware \textbf{workload representation} 
to serve as part of the input for perceiving the current tuning task~\cite{Immanuel2025@lambdatune}.
Directly feeding raw SQL statements and execution traces results in long prompts that exceed token budgets and obscure critical signals.  
Prior tuning methods rely on workload representations to guide configuration optimization~\cite{zhao22@queryformer, Kanellis22@llamatune}, which can be classified into two categories: 
(1) \textit{encoding-based} methods~\cite{zhao22@queryformer}, which map workloads to high-dimensional vectors (generalizable but often training-heavy and less interpretable), and 
(2) \textit{information-extraction} methods~\cite{Immanuel2025@lambdatune}, which extract lightweight attributes such as workload type and query templates.   
Both approaches have limitations. Encoding-based methods may over-abstract the workload, losing essential semantic details and requiring additional model training. Information-extraction methods are overly coarse, capturing only high-level characteristics, thereby constraining the LLM’s ability to reason comprehensively about the workload.

To balance expressiveness and interpretability, IDSTune employs a feature-based \textbf{workload compression} scheme consisting of two stages:
\textit{feature representation}, which encodes workloads into structured query, data, and system features (Section~\ref{sec:workload representation}); and
\textit{feature selection}, which filters relevant features for each agent (Section~\ref{sec:workload selection}).



\begin{figure}[h]
\vspace{-1em}
    \centering
\includegraphics[width=0.49\textwidth]{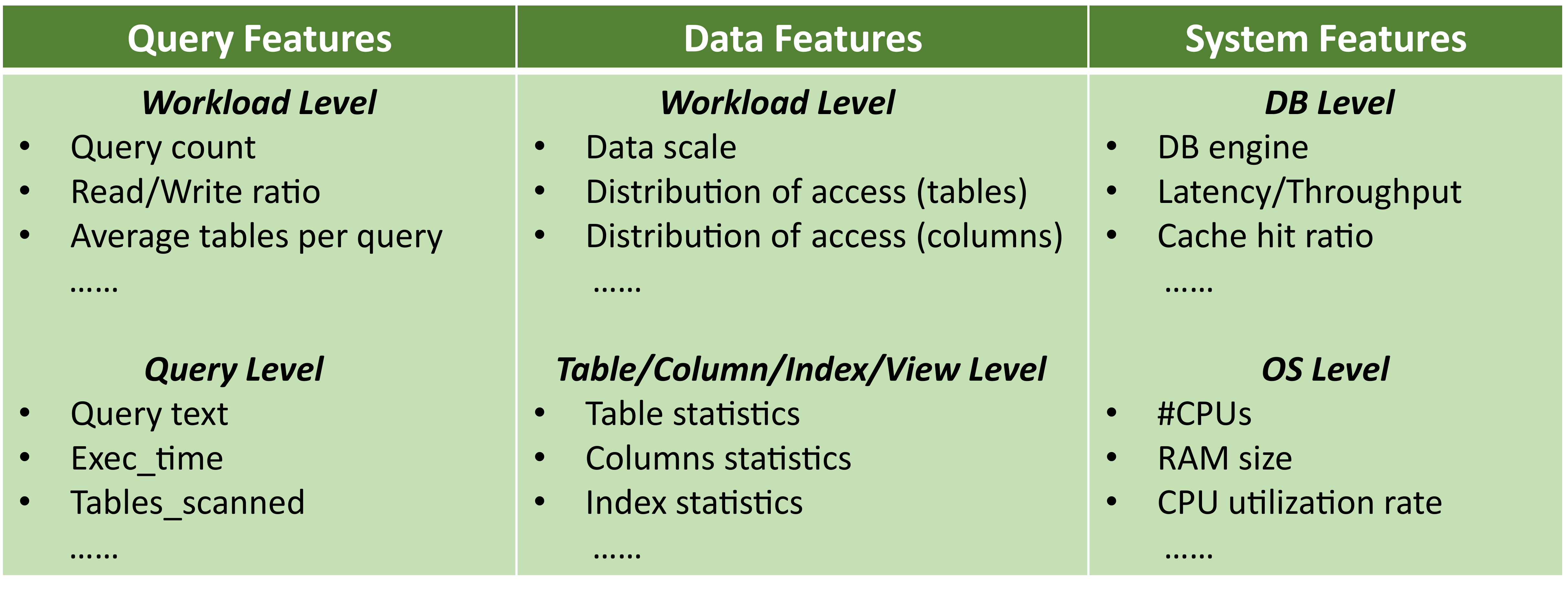}
  \vspace{-2em}
    \caption{Three categories of features in IDSTune.} 
    \label{fig:features}
    \vspace{-1em}
\end{figure}

\subsection{Workload Representation}
\label{sec:workload representation}

Motivated by how experienced DBAs carefully analyze workloads through inspecting query patterns, examining data distributions and statistics, and monitoring system status~\cite{Zhao23@tuningsurvey}, IDSTune captures three complementary categories of features, as illustrated in Fig.~\ref{fig:features}:

\begin{enumerate}[leftmargin=1.5em, itemsep=0.3em, topsep=0.3em]

  \item \textbf{Query Features} describe the characteristics of SQL statements, including statistical properties (query count, read/write ratio, average number of tables accessed) and per-query details (text, execution cost).  


    \item \textbf{Data Features} capture schema and structural information, such as tables, columns, indexes, data volume, and other properties that influence access patterns.  
  
   \item \textbf{System Features} reflect the DBMS and runtime environment, including static configurations (engine type, CPU, RAM) and dynamic metrics (CPU utilization, cache hit ratio, concurrency level) that indicate operational state.
\end{enumerate}

\begin{figure*}[t]
\vspace{-1em}
    \centering
    \includegraphics[width=0.96\textwidth]{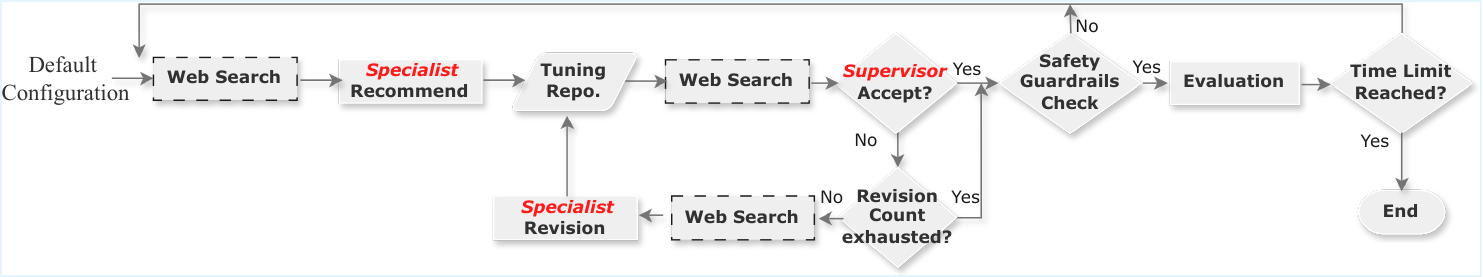}
    \vspace{-1em}
    \caption{IDSTune’s multi-agent collaborative tuning workflow.} 
    \label{fig:tuning workflow}
    \vspace{-1em}
\end{figure*}

This design combines \emph{static} signals (e.g., schema and statistics) with \emph{dynamic} runtime feedback (e.g., operator costs, resource utilization), providing a comprehensive, interpretable, and context-aware workload description.  
Unlike prior methods~\cite{Immanuel2025@lambdatune} that extract features once and rely on cost-model estimates, IDSTune maintains an evolving representation updated with real execution feedback, enabling more accurate and adaptive tuning.

As illustrated by the motivating example in Fig.~\ref{fig:performance_comparison}, when tuning PostgreSQL 15 for JOB benchmark Query 33b, increasing \texttt{work\_mem} while a related materialized view produces no rows can degrade performance due to \textbf{cardinality misestimation}, which leads the optimizer to switch from index-based plans to hash joins. 
This example highlights the importance of observing dynamic metrics, which enable IDSTune to detect such anomalies beyond static estimates.

  \vspace{-0.5em}
\subsection{Feature Selection}
\label{sec:workload selection}
After representing the workload, the next step is to identify which features are most relevant for tuning.  
Feeding all features into the LLM is inefficient, as it produces redundant prompts, increases token usage, and raises latency.  
Moreover, different specialist agents (knob, index, view) require distinct feature subsets, making heuristic selection insufficient for generalizing across workloads and tasks.

To address this, IDSTune employs an LLM-based \textbf{Selection Agent} that frames feature selection as a reasoning problem.  
\lyy{The prompt\footnote{\lyy{All complete prompts and templates in IDSTune are available at:} \url{https://github.com/intlyy/IDSTune/tree/main/prompt_template}.} is primarily composed of the following three components: }
(1) a \textbf{Task Description} that specifies the agent’s objective for the current tuning task,  
(2) \textbf{Candidate Features} are listed by name and semantics (rather than raw values) to control token cost, and  
(3) an \textbf{Output Format} that requests a compact JSON list of the selected features.  
Given this prompt, the LLM returns a minimal yet sufficient subset for the downstream specialist.
\textit{This design simulates the reasoning process of a DBA}: before starting the tuning process, the DBA observes the current workload and database environment (the candidate features), identifies the most relevant metrics for the task (the task description), and summarizes them into a concise checklist (the output format).

%% file: configuration_recommendation.tex
\vspace{-0.5em}
\section{Configuration Recommendation}
\label{sec:configuration_recommendation}
While workload compression provides concise and task-relevant context, the next challenge lies in translating it into effective configuration decisions across multiple interdependent components.  
Traditional tuning methods typically rely on a single model to optimize one configuration type in isolation.  
However, database components such as knobs, indexes, and materialized views interact in complex and often nonlinear ways, making independent optimization insufficient.  
To address this challenge, IDSTune introduces a \textbf{multi-agent collaborative framework} that coordinates specialized agents responsible for different configuration dimensions under the supervision of a \textbf{central controller}.  
IDSTune further integrates real-time DBMS feedback and external knowledge retrieval to enable adaptive and efficient tuning across all components.



\vspace{-0.5em}
\subsection{Multi-Agent Collaborative Tuning}
\label{sec:tuning_framework}
IDSTune employs a centralized multi-agent framework composed of two types of agents:
(1) \textbf{Specialist Agents}, each responsible for optimizing a specific configuration component (knobs, indexes, or materialized views); and
(2) a \textbf{Supervisor Agent}, which coordinates collaboration among specialists and manages interactions with the DBMS.
The overall workflow is illustrated in Fig.~\ref{fig:tuning workflow}.

At the beginning of each optimization round, every specialist agent independently proposes its recommendations. 
For example, the \textbf{View Specialist} suggests candidate materialized views, \lyy{the \textbf{Index Specialist} recommends appropriate index designs}, while the \textbf{Knob Specialist} recommends parameter settings. 
These outputs are consolidated into a unified optimization report covering all configuration components. 
Since specialists focus only on their respective subtasks, their recommendations may overlap or conflict (e.g., creating both an index and a materialized view on the same table). 
The Supervisor Agent inspects the report to identify potential conflicts, redundancies, or inconsistencies and instructs the corresponding specialists to refine their proposals. 
This iterative coordination continues until the supervisor deems the configuration satisfactory or the revision count is exhausted\footnote{In practice, we set the revision count to 5. While experimenting with larger values, we observed no significant performance improvements, making this setting an optimal choice for balancing LLM costs and tuning performance.}. 
Prior to execution,
the final configuration is further validated by our hybrid safety guardrails module.
The final configuration is then applied to the DBMS, feedback is collected, and the contextual information in the prompt is updated to guide the next optimization round.

In addition, all agents in IDSTune are equipped with web-search capabilities. 
Before generating their actions, agents can query external sources to retrieve relevant domain knowledge or implementation details, as described in Section~\ref{sec:search_capability}.

This collaborative tuning framework offers several key advantages:
(1) \textbf{Improved performance.} By decomposing the complex multi-component tuning task into specialized subtasks, each agent can focus on its area of expertise, leading to more effective optimization;
(2) \textbf{Reduced cost.} The modular design reduces redundant computation and promotes efficient reuse of prior knowledge; and
(3) \textbf{Transparency and interpretability.} Users are able to transparently observe how tuning recommendations are derived, along with the underlying considerations.

  \vspace{-0.6em}
\subsection{Agent Designs}
\label{sec:agent_design}
We now detail the design of the two key types of agents in IDSTune: the \textit{Specialist Agent} and the \textit{Supervisor Agent}, which together enable collaborative, efficient, and conflict-free database tuning.

\textbf{Specialist Agents.} Each specialist agent is responsible for optimizing a specific configuration component, such as knobs, indexes, or materialized views.  
All specialist agents share a unified prompt structure designed to ensure consistent reasoning and outputs across different tuning subtasks:

\begin{itemize}
[leftmargin=1.1em, itemsep=0.3em, topsep=0.3em]
    \item \textbf{Role Definition} defines the agent’s identity as an experienced DBA specializing in a particular configuration domain.
    \item \textbf{Task Description} outlines the tuning objective, i.e., to recommend database configurations that improve a specific performance metric (e.g., throughput or latency). When invoked by the supervisor agent, this section may also integrate additional constraints or revision requirements provided by the supervisor.
    \item \textbf{Workload Features} provide the subset of workload and system features selected for this agent during the \textit{Workload Compression} phase. The dynamic features are continuously updated throughout the iterative tuning process to reflect the latest DBMS state.
    \item \textbf{Current Configuration} displays the current values of the configuration parameters relevant to this agent’s domain.
    \item \textbf{External Information} represents auxiliary knowledge that assists the agent in making more informed tuning decisions. This information originates from two sources: (1) Web Retrieval: When the agent determines that the current context is insufficient for accurate tuning, it queries the web to obtain additional domain knowledge. (2) Supervisor Instructions: Guidance or revision commands issued by the Supervisor Agent, which can include specific constraints or contextual hints for refinement.
    \item \textbf{Output Format} defines the structure of the LLM’s response, ensuring consistent and parsable outputs for aggregation.
\end{itemize}

\textbf{Supervisor Agent.} The Supervisor Agent serves as the central coordinator and quality controller of the framework. It reviews consolidated tuning reports from all specialist agents to identify conflicts or redundancies. When issues arise, it provides feedback and instructs the relevant specialists to refine recommendations. This process iterates until the configuration is deemed satisfactory.

The prompt structure of the Supervisor Agent largely mirrors that of the specialists, with two key differences:
\begin{enumerate}
[leftmargin=1.5em, itemsep=0.3em, topsep=0.3em]
\item \textbf{Additional Inputs:} The Supervisor Agent receives two extra components, the \textit{tuning report} and the \textit{memory window}.   The tuning report aggregates optimization suggestions from all specialists and serves as the primary material for review, while the memory window contains historical refinement examples used as few-shot prompts to guide the supervisor’s reasoning.

\item \textbf{Output Structure:} The Supervisor Agent outputs (a) an overall decision on the current tuning report (e.g., “Accept” or “Reject”), and (b) a set of revision instructions specifying which configurations require modification and how they should be refined.  If the decision is \textit{Accept}, the configuration is applied to the DBMS; otherwise, revision commands are issued and the corresponding specialists perform targeted adjustments.
\end{enumerate}


\vspace{-1em}
\subsection{Safety Guardrails}\label{sec: safety guardrails}

While LLMs are powerful, they are prone to hallucinations. To ensure the \textbf{safety and reliability} of the tuning process, IDSTune incorporates hybrid safety guardrails that combine rule-based white-box constraints with LLM-based semantic verification.

First, the rule-based constraints layer filters out hallucinated configuration names and invalid values using constraints derived from DBMS documentation and hardware specifications. These rules are configured once per machine or DBMS with minimal manual effort. Subsequently, the LLM-based verification layer reviews the proposed configuration for logical consistency and inter-parameter dependencies, identifying subtle risks that pass static checks.
For example, setting \texttt{wal\_level} to \texttt{minimal} while keeping \texttt{max\_wal\_senders} $> 0$ may pass individual range checks but would cause a database startup failure. The LLM module detects this semantic inconsistency, whereby replication requires a higher WAL level, and rejects the configuration.

These checks are performed strictly before the configuration is applied to the database. Upon detection of an error by either layer, the specific feedback is returned to the supervisor agent to rectify the configuration. This hybrid approach ensures that only configurations that are both physically safe and semantically sound are deployed, significantly enhancing robustness.

\vspace{-0.5em}
\subsection{Search Capability}
\label{sec:search_capability}


Another key advantage of IDSTune over existing tuning frameworks lies in its integration of \textbf{external knowledge retrieval}.  
All agents can access web resources in real time to augment their reasoning with up-to-date, domain-specific information.
This design follows the principles of the ReAct framework, a widely adopted retrieval-augmented reasoning paradigm in NLP~\cite{Yao2023@ReAct}.
Before executing an action, each agent first evaluates whether the available context is sufficient for decision-making.  
If so, the agent proceeds normally; otherwise, it proactively performs a web query (via Google’s programmable search engine API~\cite{googleAPI}) to gather and summarize relevant external information.  
The retrieved content is then incorporated into the prompt as supplementary context, enabling the LLM to generate more accurate and informed recommendations.

To provide flexibility and user control, this web-search functionality can be configured in three operational modes: \textbf{Forced-On Mode:} All agents must perform a web search before every execution. \textbf{Forced-Off Mode:} The web search function remains disabled throughout the tuning process. \textbf{Auto Mode:} Each agent autonomously decides whether a web search is necessary based on its contextual confidence. As demonstrated in Section~\ref{sec:ablation_search}, these modes exhibit different trade-offs, with Auto Mode achieving the best overall balance between tuning quality and computational efficiency.
To mitigate the additional token overhead from search-augmented prompts, we omit numeric values from the \textit{Workload Features} section and retain only concise descriptions of feature names and functionalities.  
Preliminary experiments show that this strategy effectively reduces token usage and inference latency without compromising tuning performance. \lyy{In addition, we cache and reuse retrieved results across executions for efficiency and stability.}

This search-enhanced design offers several notable benefits: \textbf{(1) Efficiency}: Unlike traditional methods that pre-embed extensive auxiliary information into prompts, our agents retrieve only task-relevant knowledge on demand, reducing computational and token costs. \lyy{\textbf{(2) Adaptability}: By removing the need for manual auxiliary data management and enabling multiple operational modes, the framework adapts to diverse scenarios with minimal effort.}

%% file: evaluation.tex
\vspace{-0.5em}
\section{Experimental Evaluation}\label{sec:experiments}
This section describes the experimental setup, evaluation methodology, and results. We compare IDSTune against state-of-the-art database tuning approaches across diverse workloads and settings. \lyy{All implementations, datasets, and evaluation scripts are publicly available at: 
\url{https://github.com/intlyy/IDSTune/tree/main}.}

\vspace{-0.5em}
\subsection{Experimental Setup}
\textbf{Workloads.} 
To evaluate the performance of IDSTune, we utilize four widely adopted public benchmarks: \textbf{two OLAP benchmarks} (TPC-H~\cite{tpch} and JOB~\cite{Leis15@JOB}) and \textbf{two OLTP benchmarks} (TPC-C~\cite{tpcc} and SYSBENCH~\cite{sysbench}). Specifically, for OLAP, we use the TPC-H benchmark with a scale factor of 10, resulting in 14 GB of data and 22 complex queries. The JOB benchmark includes 113 complex queries on 9 GB of data. For OLTP, we utilize the TPC-C benchmark with 10 warehouses and 32 connections. Additionally, we employ the SYSBENCH benchmark with the OLTP-Read-Write workload, loading 50 tables with 1,000,000 rows each, resulting in a total dataset size of approximately 10 GB.

To assess the practical applicability of IDSTune, we include \textbf{three real-world benchmarks}: SDSS (16 GB), Birds (8 GB) \lyy{and Redbench}.  SDSS (seventh Data Release of Sloan Digital Sky Survey)~\cite{Abazajian09@SDSS} is an OLAP dataset containing digital astronomy data, accessible via various tools including navigation and SQL search tools.  Birds, on the other hand, is an OLTP benchmark gathered by the SQLShare~\cite{Jain16@sqlshare} project. It contains 17 tables that primarily record physiological and ecological characteristics of various bird species. \lyy{Redbench~\cite{wehrstein25@redbencd} is a recently proposed benchmark that generates trace-driven workloads derived from Amazon Redshift, enabling the evaluation under dynamic workload scenarios.}

\textbf{Hardware and Software.} Unless otherwise specified, all experiments are conducted on a machine equipped with an Intel i7-7700 processor (8 cores, 3.60 GHz) and 16 GB RAM, running PostgreSQL v15.1.  To support tuning indexes and query options in PostgreSQL, we install the HypoPG~\cite{hypopg2023} v1.4 and a patched version of pg\_hint\_plan~\cite{pghintplan} v1.6 extensions.

\textbf{Baselines.} Considering that our method supports multi- configuration optimization, we design baselines that cover single-, dual-, and triple-configuration tuning methods. To the best of our knowledge, IDSTune is the first approach capable of jointly tuning views, indexes, and knobs. Therefore, the triple-configuration baseline is constructed by combining existing methods. 

\begin{itemize} [leftmargin=1.5em, itemsep=0.3em, topsep=0.3em]
    \item \textbf{AgentTune~\cite{Li25@AgentTune} (Knob Only):} AgentTune is a state-of-the-art knob tuning framework that achieves strong optimization performance and fast convergence across diverse scenarios. It leverages LLMs to emulate a DBA’s tuning process and uses a tree-based search strategy to efficiently explore the configuration space and identify suitable knob values.
    \item \textbf{Dexter~\cite{dexter} (Index Only):} Dexter is a widely used automatic indexer for Postgres, based on HypoPG [4] and PostgreSQL optimizer’s workload costs. It works by generating hypothetical indexes and leveraging the optimizer’s cost estimates to select the index set that offers the best performance gain.
    \item \textbf{Uniview~\cite{Xu24@Uniview} (View Only):} Uniview is a unified autonomous
    materialized view management system that supports various popular databases and achieves superior performance in practical industry scenarios, leveraging greedy strategies or reinforcement learning (RL) for view selection and maintenance. 
    \item \textbf{$\lambda$-Tune~\cite{Immanuel2025@lambdatune} (Knob and Index):} $\lambda$-Tune is a framework that leverages large language models for automated database
    knob and index tuning. It generates multiple candidate knob and index configurations through iterative prompting and selects the best-performing one after evaluation. 
    \item \textbf{Proto-X~\cite{Zhang24@Holon} (Knob and Index):}  Proto-X holistically
    tunes knob and index configuration spaces. The key idea is
    to capture similarities across multiple configuration spaces, encode them into a high-dimensional representation, and synthesize “proto-actions” to navigate toward promising configurations.
    \item \textbf{HMAB~\cite{Perera22@HMAB} (Index and View):} HMAB jointly tunes indexes and MVs, using a hierarchical multi-armed bandit framework. 
    \item \textbf{Uniview + $\lambda$-Tune (Knob, Index and View):} This baseline first runs Uniview to optimize MVs, followed by $\lambda$-Tune to tune knobs and indexes. This order yields better results than the reverse, as $\lambda$-Tune requires multiple workload executions, and Uniview optimization improves the overall runtime efficiency.
    \item \textbf{Proto-X + Uniview (Knob, Index and View):} In this baseline, we replace $\lambda$-Tune in the previous setting with Proto-X and adjust the optimization order. Due to Proto-X’s candidate mechanism, the process involves frequent \texttt{EXPLAIN} and HypoPG operations, and introducing materialized views can amplify latency during online tuning. This highlights that not only the optimization methods but also their order can significantly affect both tuning effectiveness and training efficiency.
    \item \textbf{AgentTune + HMAB (Knob, Index and View):} This baseline first employs AgentTune to determine the knob configuration, followed by HMAB to optimize the index and view. The rationale behind this order is that AgentTune converges quickly and provides high-quality configurations, which accelerate HMAB’s training and improve its overall effectiveness. 
\end{itemize}

\begin{figure*}
\vspace{-0.5em}
    \centering
    \includegraphics[width=0.8\textwidth]{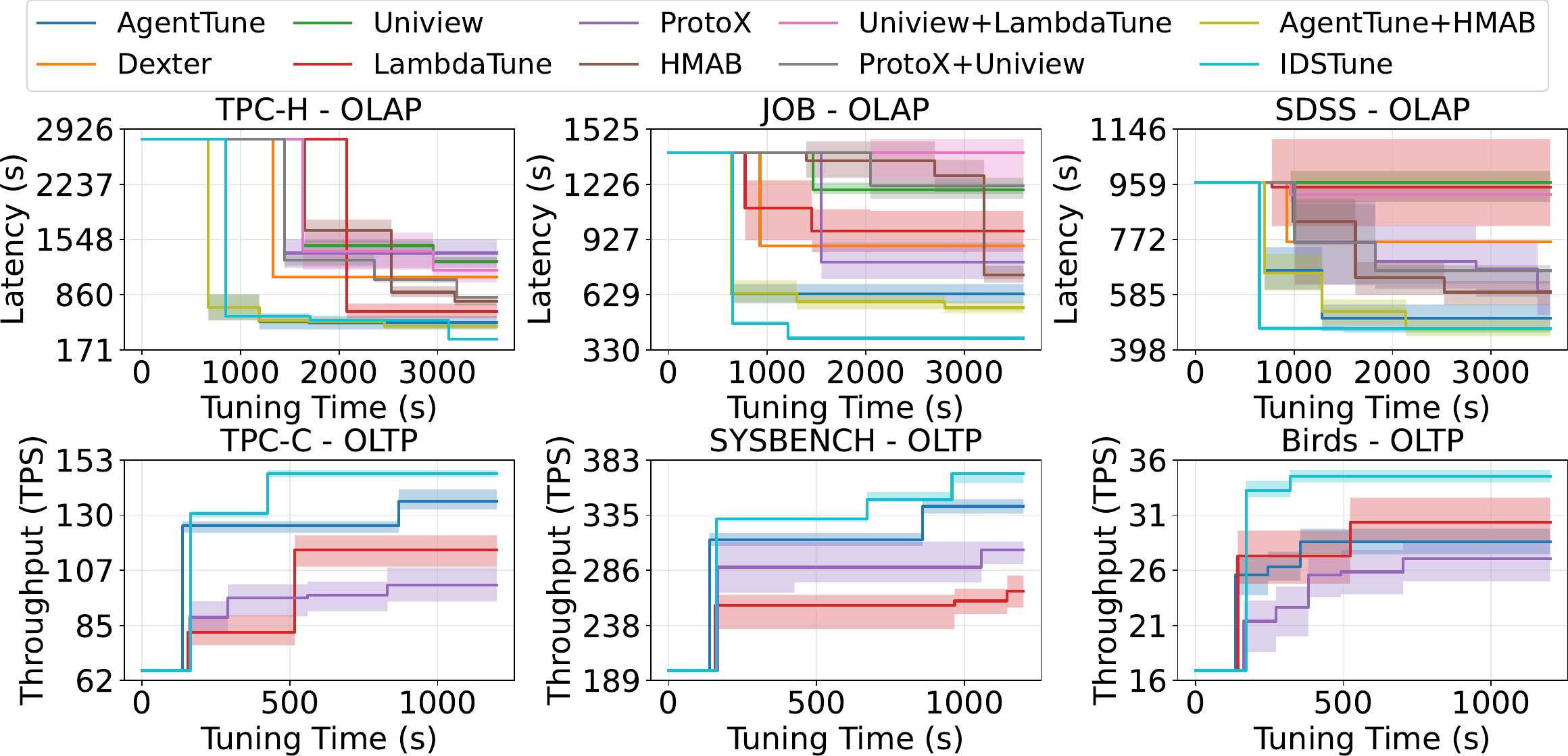}
    \caption{Performance comparison of best-found configurations throughout the tuning process across six different benchmarks. Note that at the beginning of each curve, all methods exhibit the same performance as the default configuration, since \textbf{the system needs to complete one full benchmark run to obtain valid performance feedback}. The duration before the first performance point thus corresponds to the time of configuration recommendation, application, and workload execution.} 
    \label{fig:main_result}
    \vspace{-1em}
\end{figure*}

\textbf{Tuning Settings.} \label{sec: tuning setting}
We allocate the same tuning time for all methods. In our approach, the LLM is configured with a \textbf{temperature of 0} to eliminate the influence of randomness on the results. For $\lambda$-Tune and AgentTune, the temperatures are set to 0.35 and 1, respectively, consistent with their official implementations, as both methods leverage stochasticity to generate multiple candidate configurations in a single round. We employ GPT-4.1 (this choice is justified in Section~\ref{subsubsec:ablation_llm}) as the underlying large language model for all LLM-based methods.
For UniView, which provides both Greedy-based and RL-based implementations, we adopt the Greedy-based version because our preliminary experiments show that it achieves comparable optimization performance to the RL-based version while requiring significantly less training time. For Proto-X and HMAB, we strictly follow the configurations described in their original papers. HMAB was originally implemented on Microsoft SQL Server; in our experiments, we use its PostgreSQL-compatible version.



\vspace{-0.5em}
\subsection{Main Results}
\label{sec:main_result}
The experimental results prove that our method outperforms existing state-of-the-art in both \textit{Performance}, \textit{Efficiency} and \textit{Stability}.

\noindent \textbf{IDSTune discovers superior knob configurations.} As shown in Fig.~\ref{fig:main_result}, IDSTune consistently finds the best configurations across all evaluated benchmarks. 
In the OLAP benchmark,
IDSTune consistently outperforms all baseline methods. On JOB, it achieves at least a 38.3\% and 34.7\% reduction in latency compared to non-hybrid and hybrid baseline methods, respectively, while maintaining a clear advantage in other workloads like TPC-H and SDSS ($\frac{614.87-379.25}{614.87}=38.3\%$, $\frac{580.66-379.25}{580.66}=34.7\%$).
These gains are attributed to IDSTune’s integrated optimization strategy, which considers all configuration components jointly and thereby avoids the mutual interference that often arises when tuning them independently.

In the OLTP benchmark, this trend persists. 
In terms of throughput, IDSTune outperforms baselines by at least 18.8\% on Birds, respectively ($\frac{34.53-29.07}{29.07}=18.8\%$). Similar superiority is consistently observed across TPC-C and SYSBENCH.
By contrast, several triple-configuration baselines (e.g., Uniview + $\lambda$-Tune and Proto-X + Uniview) fail on OLTP workloads. The primary reason lies in the inherent limitations of many existing tuning approaches, which are primarily designed and optimized for OLAP workloads. When applied to transactional scenarios, especially those involving materialized views, these methods may even degrade performance. 
The frequent data updates in OLTP environments can make materialized views extremely expensive to maintain, which can undermine their potential benefits.
In comparison, our framework can effectively recognize the workload type through the \textit{workload representation} and accordingly generate appropriate configuration recommendations. This adaptive capability enables IDSTune to maintain robust performance across diverse workload patterns.

\noindent \textbf{IDSTune offers the highest tuning efficiency.}  It can be observed from Fig.~\ref{fig:main_result} that IDSTune converges rapidly, achieving the optimal configuration fastest across nearly all OLAP benchmarks. For instance, on the JOB benchmark, IDSTune achieves the best-found configuration within only two optimization rounds, whereas the most competitive baseline, AgentTune+HMAB, requires significantly more effort. Compared to this leading baseline, IDSTune achieves a 56.9\% reduction in tuning time ($\frac{2803.31-1208.43}{2803.31}=56.9\%$).
For transactional workloads, although the absolute tuning time of IDSTune is not particularly outstanding, its optimization efficiency, defined as the performance improvement ratio per unit time, remains superior to nearly all existing approaches.

This significant efficiency advantage can be attributed to two main factors:
(1) In-framework iterative refinement. The multi-agent architecture of IDSTune enables partial “in-context iteration” during the recommendation stage. Through communication among agents, candidate configurations are self-assessed and refined before execution, resulting in higher-quality recommendations. In comparison, existing approaches, whether ML-based or LLM-based, almost entirely rely on actual database executions to evaluate configuration quality, a process that is both time-consuming and costly. For OLAP workloads, the workload execution time substantially exceeds the algorithmic recommendation time. Hence, allocating a longer and more deliberate recommendation phase to achieve higher-quality configurations is justified. Further analysis is provided in Section~\ref{sec:cost_analysis}. (2) Parallel optimization of multiple configuration components. IDSTune jointly optimizes all configuration types (knobs, indexes, and views) in parallel, whereas existing multi-component baselines such as UniView + $\lambda$-Tune, Proto-X + UniView, and AgentTune + HMAB adopt sequential optimization strategies, leading to significantly higher latency.

\vpara{IDSTune generalizes well to real-world workloads.}  
When deployed in a production environment, our method consistently delivered strong performance. Under the OLAP workload (SDSS), IDSTune successfully reduces the latency from 965.48 s (default configuration) to 465.43 s, outperforming all baselines by an average of 32.7\% ($\frac{783.56-527.39}{783.56}=32.7\%$).
For OLTP workload (Birds),  IDSTune also achieves the highest throughput, exceeding the average performance of other baselines by $1.6\times$ times ($\frac{34.53}{21.03}=1.6$).
These results demonstrate that IDSTune can effectively generalize to real-world tuning tasks, even in real-world production environments that LLMs have never encountered before, highlighting its strong transferability and practical utility.

\vpara{\lyy{IDSTune demonstrates optimal stability across multiple repeated runs.}} 
\lyy{To rigorously evaluate the robustness and reproducibility of all tuning algorithms, we conduct five independent tuning sessions per method and report the median along with the interquartile range (Q1–Q3) of the best observed performances. The shaded regions in Figure~
\ref{fig:main_result} illustrate the performance variability of each method over the tuning process.}

\lyy{Across all benchmarks, IDSTune exhibits the narrowest interquartile ranges, indicating minimal performance fluctuation and strong robustness compared to baselines. This stability can be attributed to two key design choices. \ding{202} The integrated LLM is configured with a temperature of 0, effectively eliminating stochasticity during the configuration generation process. \ding{203} Our search framework employs a history-aware caching mechanism that records previously evaluated configurations and trajectories, thereby avoiding redundant exploration and ensuring deterministic behavior.}

\vspace{-0.5em}
\begin{figure}
    \centering
    \includegraphics[width=0.46\textwidth]{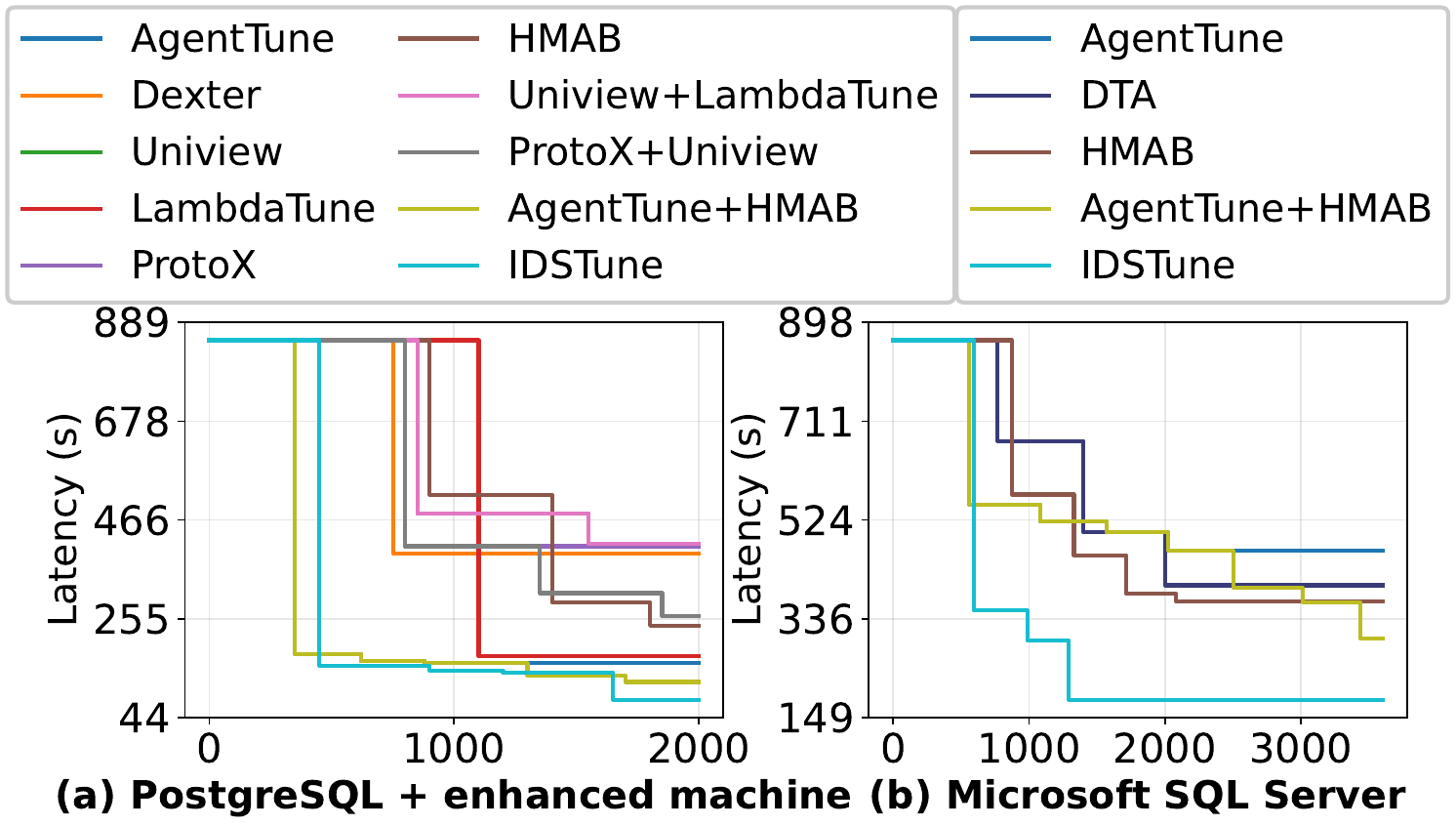}
      \vspace{-0.5em}
    \caption{\lyy{Experiments on different: (a) hardware, and  (b) database engine. }} 
    \label{fig:scalability}
    \vspace{-1em}
\end{figure}

\input{tables/result_layout}

\subsection{Scalability Study}\label{sec:scalability}
\subsubsection{Database Scaling}
We now evaluate the scalability of IDSTune by examining its performance across different database scales. Specifically, we conduct experiments on the TPC-H benchmark with four data sizes: 1 GB, 5 GB, 10 GB, and 20 GB. The results are summarized in Table~\ref{tab:db scaling}.

As database scale increases, absolute performance degrades for both default and optimized configurations, as expected due to higher data volume and execution complexity. Yet, IDSTune consistently delivers substantial improvements of $7-9\times$ over the default configuration across all scales, demonstrating robust performance under scaling (($\frac{78.19}{11.73}=7, \frac{6013.04}{635.95}=9$)).

Importantly, the overhead of IDSTune, in terms of both token consumption and runtime, remains largely constant and does not grow with the database size. This is because our workload compression mechanism is based on feature extraction rather than directly encoding raw workloads. Consequently, for larger and more complex workloads with long execution times (e.g., hours per run), the additional tuning overhead introduced by IDSTune becomes negligible and is well justified by the resulting performance gains.


\vspace{-0.5em}
\subsubsection{\lyy{Hardware Scaling}}
\lyy{As detailed in Section~\ref{sec: tuning setting}, our primary evaluations are conducted on a server with an 8-core CPU and 16 GB of RAM. To investigate the impact of hardware on tuning performance, we deploy the database on a machine equipped with 40 cores (Intel Xeon Gold 5118 @ 2.30GHz) and 256 GB of RAM. The experimental outcomes are presented in Figure~\ref{fig:scalability} (a).}

\lyy{We observe that the relative effectiveness of different methods becomes less pronounced at larger scales. Approaches that optimize knobs achieve significant performance gains, primarily because, under abundant resources such as memory, suboptimal configurations become the performance bottleneck compared to physical structures like indexes. Meanwhile, IDSTune consistently outperforms all baselines and achieves the lowest final latency, demonstrating its superior ability to adapt to \textbf{diverse hardware} environments.}

\vspace{-0.5em}
\subsubsection{\lyy{Database Engine Generalization}}

\lyy{
To evaluate cross-engine generalization, we replace PostgreSQL with Microsoft SQL Server \cite{mssqlserver} while keeping all other experimental settings unchanged. Due to limited support of some baselines on SQL Server, we additionally include Database Tuning Advisor (DTA)~\cite{DTA}, a competitive physical design tuning approach specifically designed for this database engine, as a strong baseline.
}

\lyy{
As shown in Figure ~\ref{fig:scalability} (b), IDSTune consistently maintains superior performance on SQL Server, outperforming all applicable baselines. While some methods suffer from degraded effectiveness due to their reliance on database-specific features, IDSTune remains robust by leveraging its search capability to retrieve relevant knowledge and its LLM-driven coordination across tuning actions. Compared to DTA, which focuses solely on physical design optimization, IDSTune delivers better overall performance by jointly optimizing multiple configuration dimensions. Furthermore, IDSTune is easy to deploy in practice, as it adopts a training-free design that avoids additional preparation overhead. These results highlight the strong \textbf{cross-engine generalization ability} of IDSTune.
}

\vspace{-0.5em}
\subsection{Robustness Against Drift}
\lyy{In real-world deployments, the DBMS environment under tuning may vary due to data drift, query drift, or workload changes. Given this, we now analyze whether IDSTune can still effectively guide the database toward promising configurations when its memory becomes outdated. We first examine data and query drift, followed by a real-world evolving workload trace.  

For combined baselines, components are executed sequentially under static workloads. However, handling drift requires re-activating previously completed processes, which most methods do not support. Therefore, they are not considered in this section.
}
\subsubsection{Data Drift}
We construct a progressive data drift scenario based on the JOB benchmark. Specifically, we partition the dataset using the \texttt{production\_year} attribute. The initial state contains only historical data (years < 2005). We then introduced new data in 20\% increments (Stages 20\% to 100\%), creating five distinct drift events. Each stage was allocated the same tuning budget of time. We utilized the original queries from the JOB benchmark. 

Figure \ref{fig:drift} reports the performance trajectory across the six stages. We observe that IDSTune outperforms baselines in two key aspects. First, it exhibits minimal performance degradation at the onset of data drift. Unlike methods that rely solely on fragile physical design structures (indexes and materialized views) which are prone to invalidation during distribution shifts, IDSTune incorporates knob tuning as a stabilizing factor. Knobs generally retain their efficacy even when data distribution changes, providing a performance buffer. Second, IDSTune demonstrates rapid recovery following drift events. This is attributed to our adaptive feature extraction mechanism. By refreshing the features prior to each optimization round, IDSTune timely detects the underlying distribution shifts, enabling the agent to adjust its search strategy immediately.

\begin{figure}
\vspace{-0.5em}
    \centering
    \includegraphics[width=0.46\textwidth]{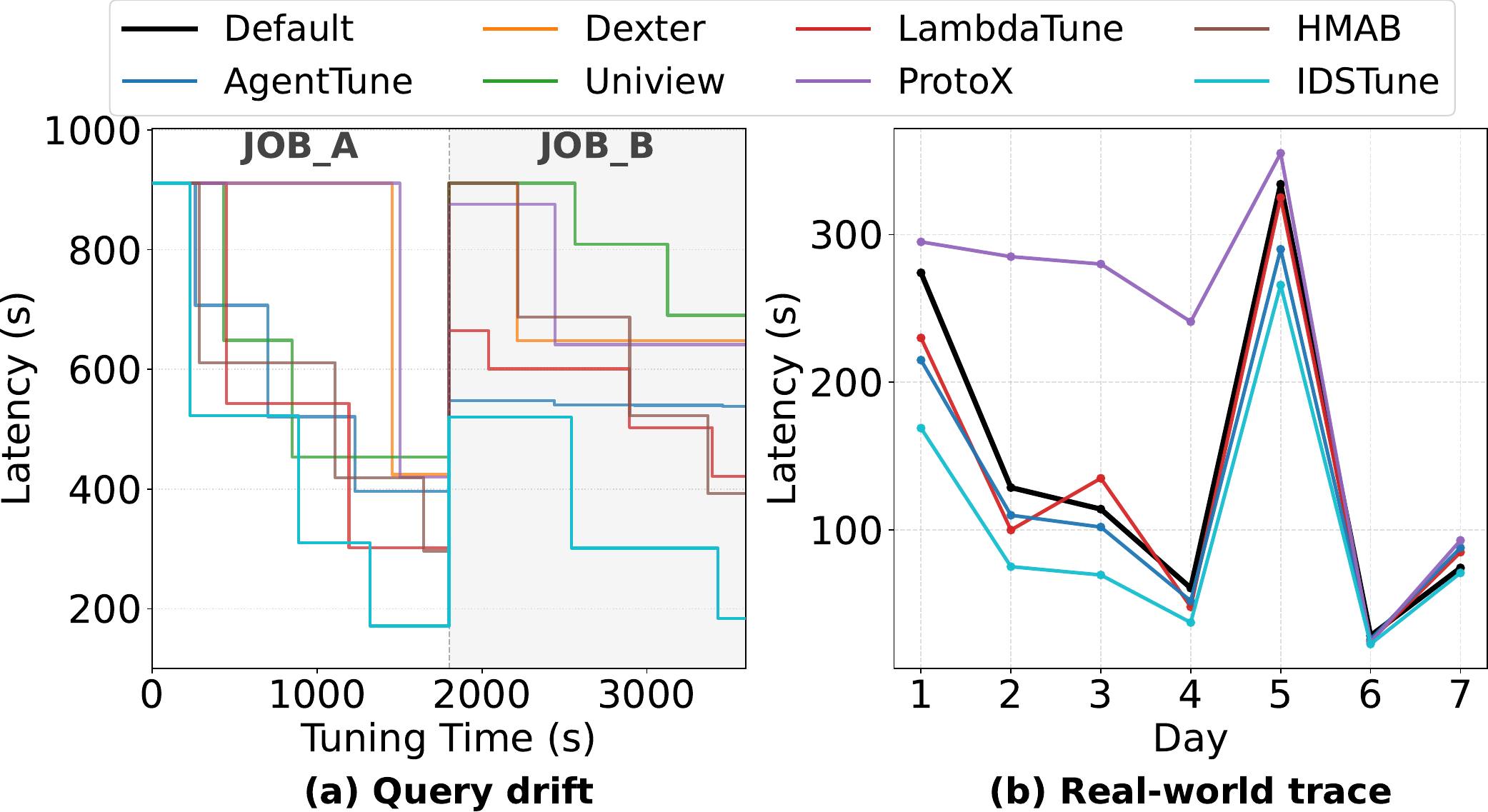}
      \vspace{-1em}
    \caption{\lyy{Experiments on: (a) query drift, and  (b) real-world trace. }} 
    \label{fig:query drift}
    \vspace{-2em}
\end{figure}

\subsubsection{\lyy{Query Drift}}
\lyy{To evaluate IDSTune’s resilience to schema-level query drift, we divide the JOB workload into two disjoint sets. Specifically, based on the 33 query templates in JOB, we randomly split the templates and their derived queries into two subsets, resulting in \textbf{JOB\_A} (16 templates, 48 queries) and \textbf{JOB\_B} (15 templates, 65 queries). This partition ensures minimal overlap between the two subsets, thereby reducing the extent to which methods can benefit from previously observed query patterns.}

\lyy{Figure~\ref{fig:query drift} (a) presents the results. We observe that methods incorporating knob tuning are generally less affected by query drift and recover faster than those focusing solely on physical design optimization. This is mainly because knob configurations capture system-level performance characteristics that are less tightly coupled with specific query patterns. On top of this, IDSTune further outperforms all baselines. Its stronger resilience to drift comes from its richer feature representation, which incorporates not only query features but also data features that remain stable under query drift. Moreover, IDSTune adapts more quickly because its features are dynamically updated after each execution, allowing it to respond promptly to newly arrived queries.}

\vspace{-0.8em}
\subsubsection{\lyy{Real Workload Trace}}
\lyy{
Compared with classical benchmarks such as TPC-H and JOB, real-world production workloads are typically more write-intensive and evolve over time, exhibiting both data and query drift~\cite{Alexander24@TPCisnotenough}. To evaluate the effectiveness of tuning methods under such realistic conditions, we use Redbench~\cite{wehrstein25@redbencd} to emulate real-world workload trace. 
Specifically, we construct a one-week workload trace based on JOB templates, consisting of SELECT  and diverse DML operations. The tuning methods are required to optimize the database at the end of each day. This setup implies that the workload of the next day is never observed during tuning, posing a continuous adaptation challenge.}

\lyy{The results are shown in Figure~\ref{fig:query drift} (b), where the black curve represents the performance under the default configuration. IDSTune achieves an average improvement of approximately 29.9\% over the default configuration ($\frac{144.93-101.54}{144.93}=29.9$). Importantly, in more complex real-world scenarios, the performance gains become more pronounced, while the associated optimization cost remains nearly unchanged. We also observe that, under such continuously evolving workloads, existing methods are prone to negative optimization, especially those relying heavily on historical data. In contrast, LLM-based methods, such as LambdaTune and AgentTune, exhibit more stable performance. IDSTune further distinguishes itself as the \textbf{only} method that consistently achieves positive improvements throughout the entire trace, which is particularly critical for real-world scenarios where performance regressions are unacceptable. This advantage stems from its rich feature representation and multi-agent coordination mechanism, which enable robust adaptation under dynamic workload conditions. }

\vspace{-0.5em}
\subsection{Ablation Studies}
To further evaluate the effectiveness of each component in our framework, we conduct comprehensive ablation studies throughout the entire tuning process. In addition, we examine the impact of different LLMs on tuning performance. Unless otherwise specified, all ablation experiments are conducted on JOB benchmark.

\subsubsection{Ablation Study on the Workload Compression}
First, we conduct ablation studies to evaluate the effectiveness of our workload compression module. Specifically, we compare three different workload representation strategies: (1) Workload representation + selection, i.e., the method used in IDSTune; (2) Workload representation only, where the workload is embedded without feature selection; (3) No 
workload, where the workload compression module is removed, and only the most basic hardware and database information can be accessed.

The results are shown in Table~\ref{tab:ablation_compression}. Our workload compression approach achieves the best overall performance, as IDSTune provides a comprehensive and fine-grained workload representation that allows the LLM to more effectively perceive and reason about query patterns. However, this does not imply that simply increasing the amount of information always improves results. As demonstrated by the "Representation Only" variant, providing the full unfiltered feature set can actually degrade performance, since not all information is useful. Even when the workload compression module is completely removed, our method still achieves a noticeable performance improvement over the default configuration, demonstrating that IDSTune remains effective even under extreme conditions.

\subsubsection{Effectiveness of Multi-Agent Collaborative Tuning Framework}

To validate the effectiveness of our collaborative multi-agent architecture, we compare IDSTune with two simplified designs:
(1) \textit{Multi-agent non-collaborative}, where the supervisor is removed and each specialist agent (knob, index, view) operates independently without inter-agent communication; and
(2) \textit{Single-Agent}, which merges all agents into a single LLM call that recommends all configuration types at once.

\input{tables/ablation_multi-agent}
The results are reported in Table~\ref{tab:ablation_framework}. We make three key observations.
(1) Both the multi-agent non-collaborative variant and the single-agent baseline consistently underperform IDSTune across all workloads. This performance gap highlights the importance of explicit collaboration among specialized agents, especially in complex tuning scenarios where different configuration types interact and jointly affect system performance. By enabling structured communication and iterative coordination, IDSTune avoids suboptimal decisions that arise when configuration dimensions are optimized in isolation or entangled within a single monolithic reasoning process.
(2) IDSTune incurs higher token consumption and tuning runtime than the simplified baselines. This overhead mainly stems from maintaining multiple specialized agents and enabling inter-agent communication and iterative coordination. Importantly, such overhead is largely fixed and does not scale with workload size or execution complexity, as discussed in Section~\ref{sec:scalability}. This ensures that, for complex workloads with long execution times (e.g., hours per run), the additional tuning cost remains negligible relative to the performance gains.

\begin{figure}[!t]
  \centering
  \begin{subfigure}{0.47\columnwidth}
    \centering
    \includegraphics[width=\textwidth]{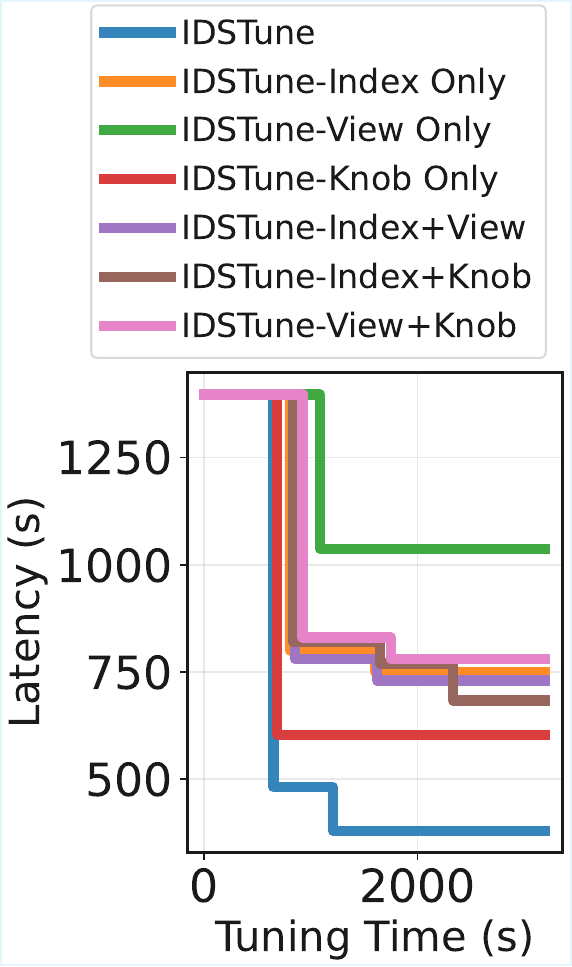}
    \caption{Agent Composition}
  \end{subfigure}
  \hfill
  \begin{subfigure}{0.51\columnwidth}
    \centering
    \includegraphics[width=\textwidth]{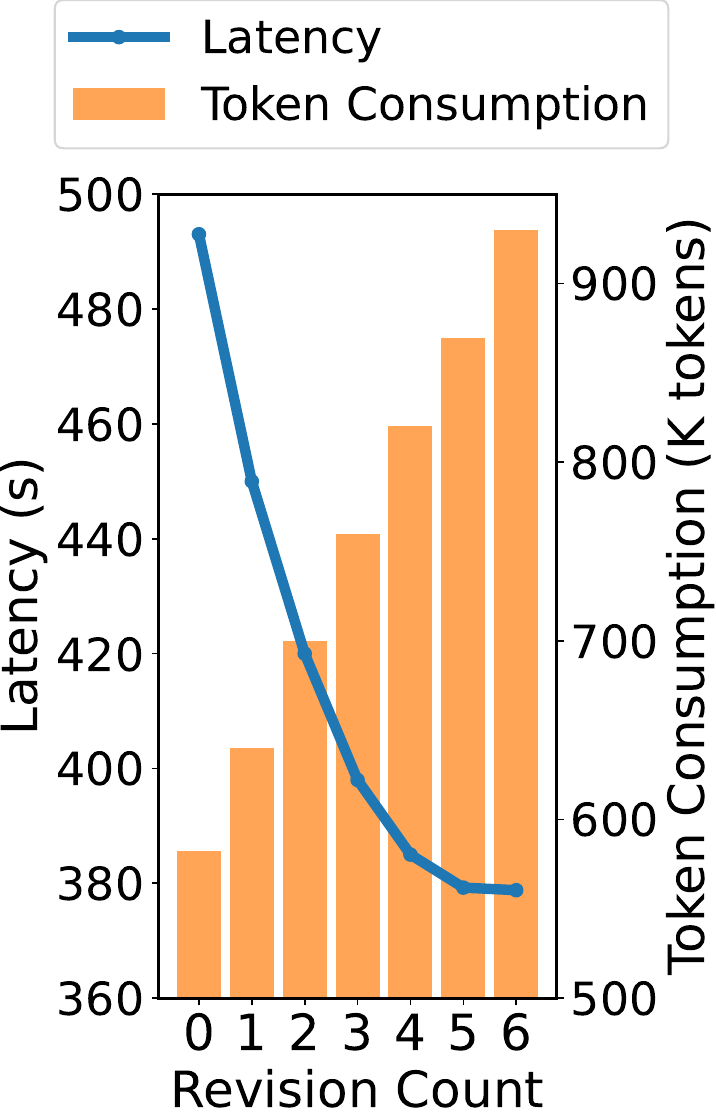}
    \caption{Revision Count}
  \end{subfigure}
  \vspace{-0.6em}
  \caption{Ablation studies on: (a) agent composition, and (b) revision count.}
  \label{fig:combined1}
  \vspace{-2.1em}
\end{figure}

\subsubsection{Ablation study on Agent Composition} 
To understand how different types of tuning agents complement each other and contribute to overall tuning effectiveness, we compare the following variants: (1) IDSTune-Index/View/Knob Only, where only one type of agent is retained; (2) IDSTune-Knob \& Index, keeping only knob and index agents; (3) IDSTune-Knob \& View, keeping knob and materialized view agents; (4) IDSTune-View \& Index, keeping materialized view and index agents.


Fig.~\ref{fig:combined1}(a) presents the results of the ablation study. 
We observe that applying individual components in isolation or combining them in pairs can still lead to noticeable performance improvements over the default configuration, indicating that each specialized agent contributes meaningful optimization capability. This also demonstrates the flexibility of IDSTune, as the framework can be adapted to different practical requirements by selectively enabling 
specific agents. However, no single agent or agent pair can consistently match the performance of the full collaborative framework, highlighting the complementary nature of different tuning dimensions and the importance of holistic coordination among agents.

\subsubsection{Ablation Study on Revision Count}\label{subsubsec:ablation_revision_count}
We now study the impact of the revision count, which controls the maximum number of iterative refinements allowed when agents fail to reach consensus during collaborative tuning. Intuitively, a larger revision count enables more thorough coordination among agents, but also incurs higher LLM inference costs. This experiment aims to quantify this trade-off and identify a practical setting.


We vary the revision count from 0 to 6, keeping all other configurations unchanged. Fig.~\ref{fig:combined1}(b) reports tuning latency and token consumption under different revision counts. Increasing the revision count initially improves latency, as additional iterations allow agents to refine decisions and resolve inconsistencies. However, gains diminish after five revisions, indicating convergence. In contrast, token consumption grows steadily with the revision count due to extra LLM calls. Consequently, increasing revisions beyond five yields marginal performance improvements at a higher cost. Based on this, we set the revision count to five, achieving a balance between tuning effectiveness and LLM overhead.

\subsubsection{Ablation study on safety guardrails}\label{subsubsec:ablation_safety}

\input{tables/ablation_safety}
Our safety guardrails integrate rule-based constraints and LLM-based verification to synergistically guarantee tuning reliability and prevent invalid configurations. To evaluate the specific contribution of each component, we conducted an ablation study comparing the following four variants: (1) IDSTune: The complete framework with full safety mechanisms; (2) w/o rule-based constraints: The variant with the static constraints removed; (3) w/o LLM-based verification: The variant with the semantic verification disabled; (4) w/o safety guardrails: The variant where the entire safety mechanism is excluded.

The results are summarized in Table~\ref{tab:ablation_safety}. \lyy{False Negative indicates a failure to intercept, resulting in unsafe configurations being applied, whereas False Positive refers to incorrectly flagging a valid configuration as unsafe. }As observed, the complete IDSTune system maintained a zero invalid configuration rate throughout the process, demonstrating the robustness and reliability of our approach. In contrast, all variants lacking specific safety components generated unsafe configurations.  Consequently, without guardrails to intercept these unreasonable configurations, the optimization process is disrupted, negatively impacting the final performance.
Another noteworthy observation is regarding token consumption. Although the LLM-based verification module introduces additional LLM calls \lyy{and the risk of false positives}, removing it surprisingly resulted in increased total token usage. This is primarily because the prompts used for verification are concise and executed only prior to deployment, incurring minimal overhead. Conversely, the absence of verification leads to wasted iterations on invalid configurations. The cost of these failed trials—which require re-generating contexts and restarting the dialogue—significantly outweighs the marginal token cost of the verification prompts.

\subsubsection{Ablation Study on the Search Mechanism}
\label{sec:ablation_search}
\input{tables/ablation_search}
Our framework allows agents to retrieve external knowledge from the Web and provides three operational modes:
(1) \textit{Forced-On}, where all agents perform a web search before every execution;
(2) \textit{Auto}, where each agent decides autonomously whether a web search is needed based on contextual confidence; and
(3) \textit{Forced-Off}, where web search is disabled throughout the tuning process.

Table~\ref{tab:ablation_search} presents the performance, time cost, and token consumption under different modes. We can observe that agents generally benefit from search, as it allows them to acquire up-to-date or domain-specific knowledge during tuning. However, this improvement comes at a cost, i.e., longer execution time or higher token usage.
Therefore, for users who prioritize optimization performance, the Forced-On mode is preferable, whereas the Forced-Off mode is more suitable for time- or token-sensitive scenarios. Overall, the Auto Mode achieves the best trade-off between performance and efficiency, and thus serves as the default setting in IDSTune.

\subsection{\lyy{Sensitivity Experiments}}
\lyy{To better understand the sensitivity of IDSTune to key factors, We next analyze its behavior under different settings. We begin with an ablation study on memory budget in Section~\ref{sec: ablation on memory budgets}, followed by analyses of the effects of tuning time and LLMs in Section~\ref{sec: effect of tuning time} and  Section~\ref{subsubsec:ablation_llm}, respectively.}
\begin{figure}
\vspace{-0.5em}
    \centering
    \includegraphics[width=0.46\textwidth]{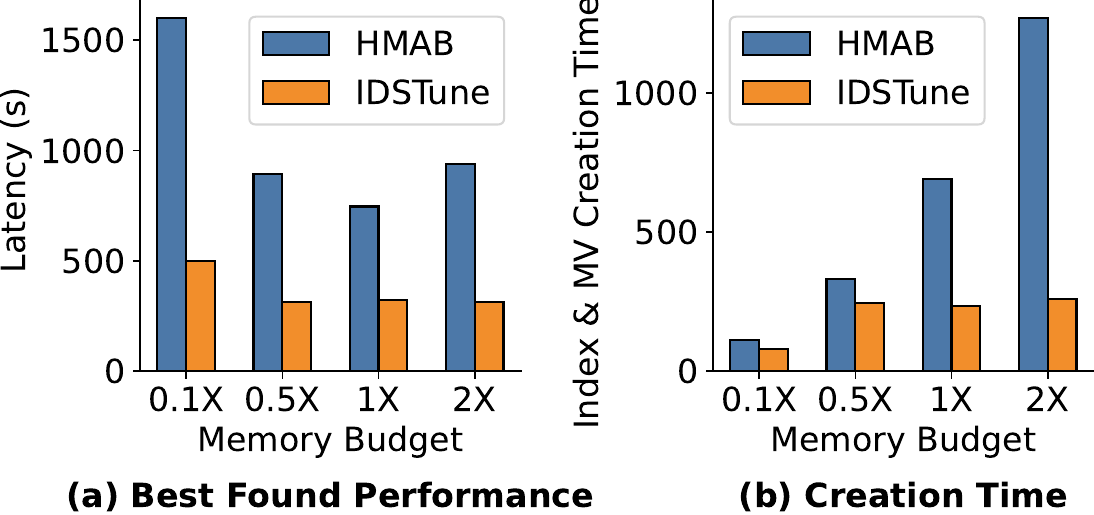}
      \vspace{-1em}
    \caption{\lyy{Impact of memory budget on: a) performance and b) creation time. }} 
    \label{fig: memory budget}
    \vspace{-1.5em}
\end{figure}

\subsubsection{\lyy{IDSTune’s Performance under Different Memory Budgets}} \label{sec: ablation on memory budgets}
\lyy{This section presents experiments across four different memory budgets, 0.1X, 0.5X, 1X (approximately equal to the data size), and 2X, under TPC-H benchmark, to understand the memory budget’s impact on solution fitness. Among all baselines, only our method and HMAB can actively control the memory budget. 
As shown in Figure~\ref{fig: memory budget}(a), both IDSTune and HMAB perform poorly under low memory budgets. As the allocated memory increases, IDSTune steadily improves and eventually converges, whereas HMAB exhibits slight performance degradation after convergence. This behavior is mainly attributed to HMAB’s search mechanism, which tends to aggressively explore a large number of physical design candidates. When the search space becomes excessively large, such exploration incurs substantial overhead and can even hinder performance. 
In contrast, IDSTune’s LLM-based multi-agent framework enables more disciplined utilization of the allocated memory, leading to more stable performance. This is further validated in Figure~\ref{fig: memory budget}(b), where IDSTune’s physical design creation time does not continuously 
increase with the memory budget.}

\begin{figure}
    \centering
    \includegraphics[width=0.48\textwidth]{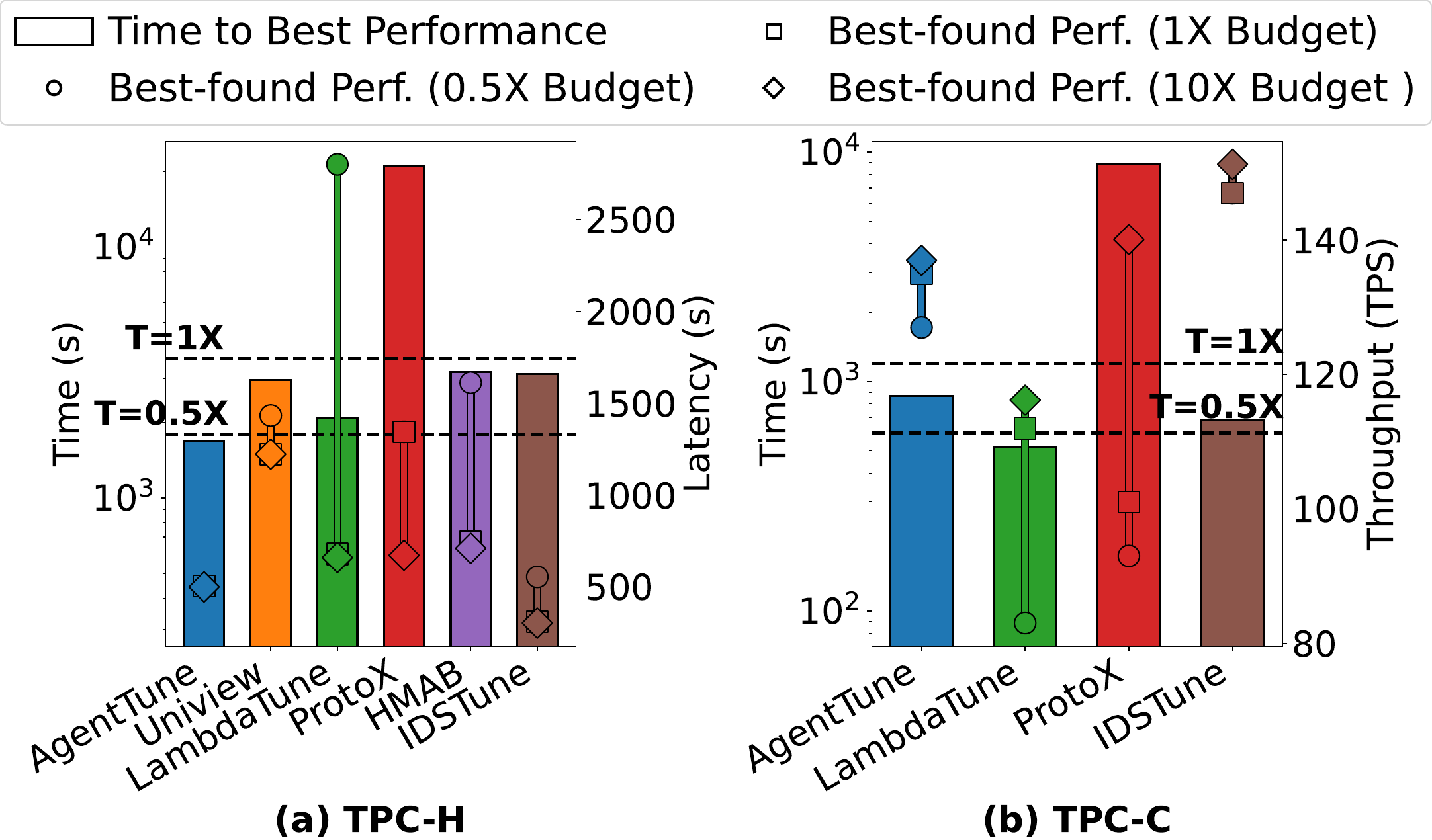}
      \vspace{-2em}
    \caption{\lyy{Impact of time budget on performance.}} 
    \label{fig: memory budget}
    \vspace{-0.5em}
\end{figure}

\vspace{-0.5em}
\subsubsection{\lyy{Effect of Tuning Time}} \label{sec: effect of tuning time}
\lyy{
As discussed in Section~\ref{sec: tuning setting}, we allocate the same tuning time for all methods (3600s for OLAP workloads and 1200s for OLTP workloads).
To examine the impact of the time budget, we evaluate each method under 0.5$\times$ and 10$\times$ the original setting on TPC-H (OLAP) and TPC-C (OLTP). As shown in Figure~\ref{fig: memory budget}, the bar chart reports the time required for each method to reach its best-found configuration under the extended budget, while the markers illustrate the best performance achieved under different time budgets. To account for inherent performance variability in database systems, we adopt a 5\% improvement threshold when determining whether a new configuration is better. We observe that our method consistently achieves the best performance across all time budgets. Moreover, for most methods, the original time budget is sufficient to reach near-optimal performance. The only exception is ProtoX, which appears to benefit from longer tuning time, likely due to its reinforcement learning nature requiring more training iterations. In contrast, for LLM-based methods such as LambdaTune, excessively long tuning durations may incur additional computational cost without proportional performance gains. Thus, from a holistic perspective that considers both effectiveness and efficiency, the current time settings strike a reasonable balance.}

\begin{figure}[t]
  \centering
  \begin{minipage}{0.46\columnwidth}
    \centering
    \includegraphics[width=\textwidth]{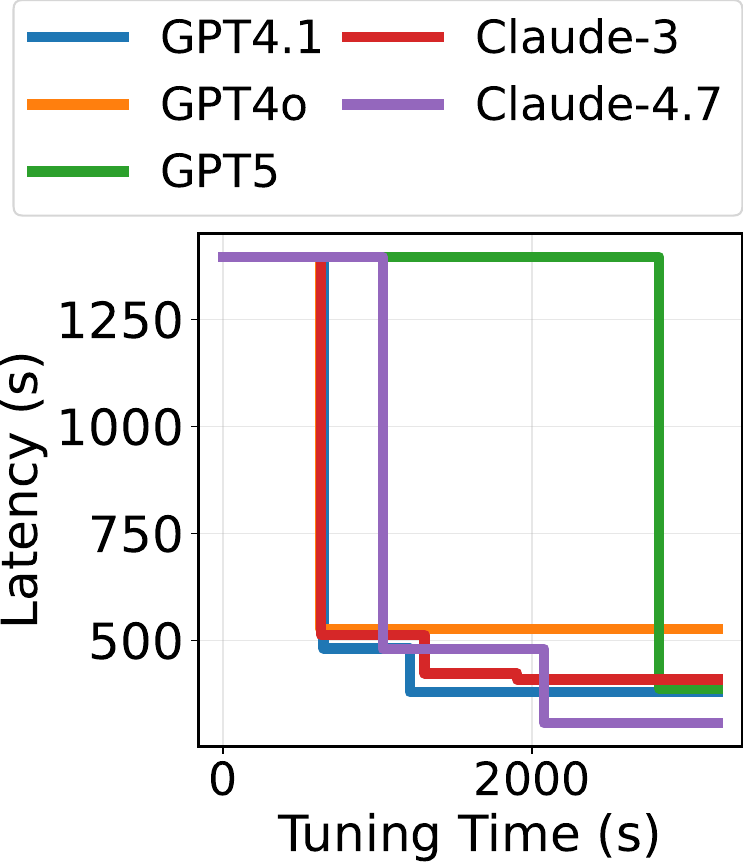}
    \vspace{-2em}
    \caption{Different LLMs.}
    \label{fig:ablation_llm}
  \end{minipage}
  \hfill
  \begin{minipage}{0.48\columnwidth}
    \centering
    \includegraphics[width=\textwidth]{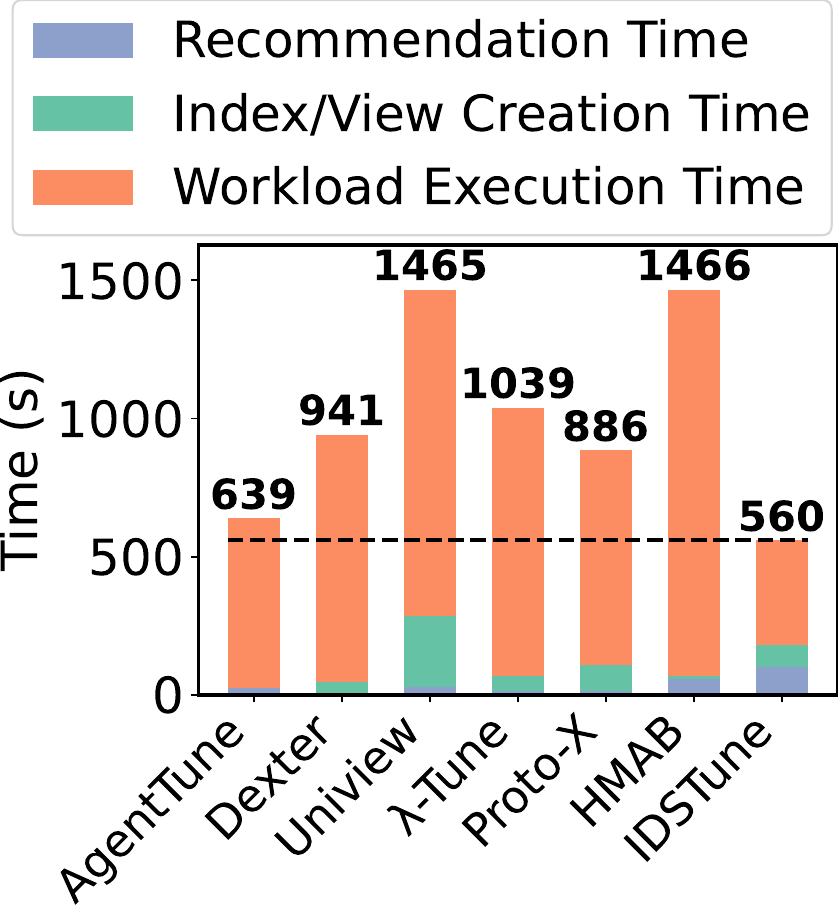}
    \vspace{-2em}
    \caption{Time breakdown.}
    \label{fig:cost}
  \end{minipage}
  \vspace{-1em}
\end{figure}

\subsubsection{Effect of Different LLMs}\label{subsubsec:ablation_llm}
We analyze the effect of LLMs on tuning performance. Specifically,
IDSTune is tested with  GPT-5~\cite{gpt5}, GPT-4.1~\cite{openai2025@gpt4.1}, GPT-4o~\cite{openai2024@gpt4o}, Claude-3~\cite{Claude_3} and Claude-4.7~\cite{Claude4.7} on JOB benchmark. As shown in Fig.~\ref{fig:ablation_llm}, all variants achieve consistent and strong performance, indicating that the effectiveness does not rely on any specific LLM.
\lyy{We observe that more powerful models, such as the recently released Claude-4.7, can further improve performance. While it achieves a 19.2\% improvement over our primary model, it is around 4$\times$ more expensive and has a longer runtime. 
Nevertheless, we expect that such models will become more cost-efficient and faster over time. This trend suggests that our approach can naturally benefit from future advances in LLMs.}


%
\vspace{-0.5em}
\subsection{Cost Analysis}
\label{sec:cost_analysis}



There are two main types of overhead in DBMS knob tuning:  (1) initial profiling overhead and (2) runtime overhead. Initial profiling overhead refers to the time required to collect training data or pre-train models before tuning begins (e.g., RL model in Uniview requires tens of hours of training for cold start; AgentTune needs to prepare and process relevant information in advance before tuning [26]). Runtime overhead is the
time the tuner takes to recommend a new configuration for evaluation.

IDSTune offers out-of-the-box flexibility with minimal initial profiling overhead compared to previous methods. We evaluate the runtime overhead by presenting the number of tokens consumed, monetary costs, and time required using GPT-4.1. In the main experiments (Section~\ref{sec:main_result}), IDSTune incurs 2977.46 K tokens, with a total cost of 13.82 USD. Furthermore, we compare the time consumption of different methods in a single tuning process, which can be divided into three stages: algorithm recommendation time, index/view creation time, and workload execution time, as shown in Fig.~\ref{fig:cost}. We observe that the configuration recommendation time is generally negligible compared with the workload replay time and the physical design creation time. Compared with other methods, IDSTune incurs a relatively longer algorithm recommendation time, mainly due to the iterative communication among multiple agents and the refinement of the tuning report. However, this additional cost leads to the shortest workload execution time (i.e., the best achieved database performance) and thus the shortest overall tuning time. In summary, IDSTune’s LLM-related costs are feasible and practical, making it suitable for real-world large-scale deployment.



%% file: tables/result_layout.tex
\begin{figure*}[t]
  \centering
  \begin{minipage}[t]{0.6\textwidth} 
    \centering
    \vspace{0pt}
    \includegraphics[width=\textwidth]{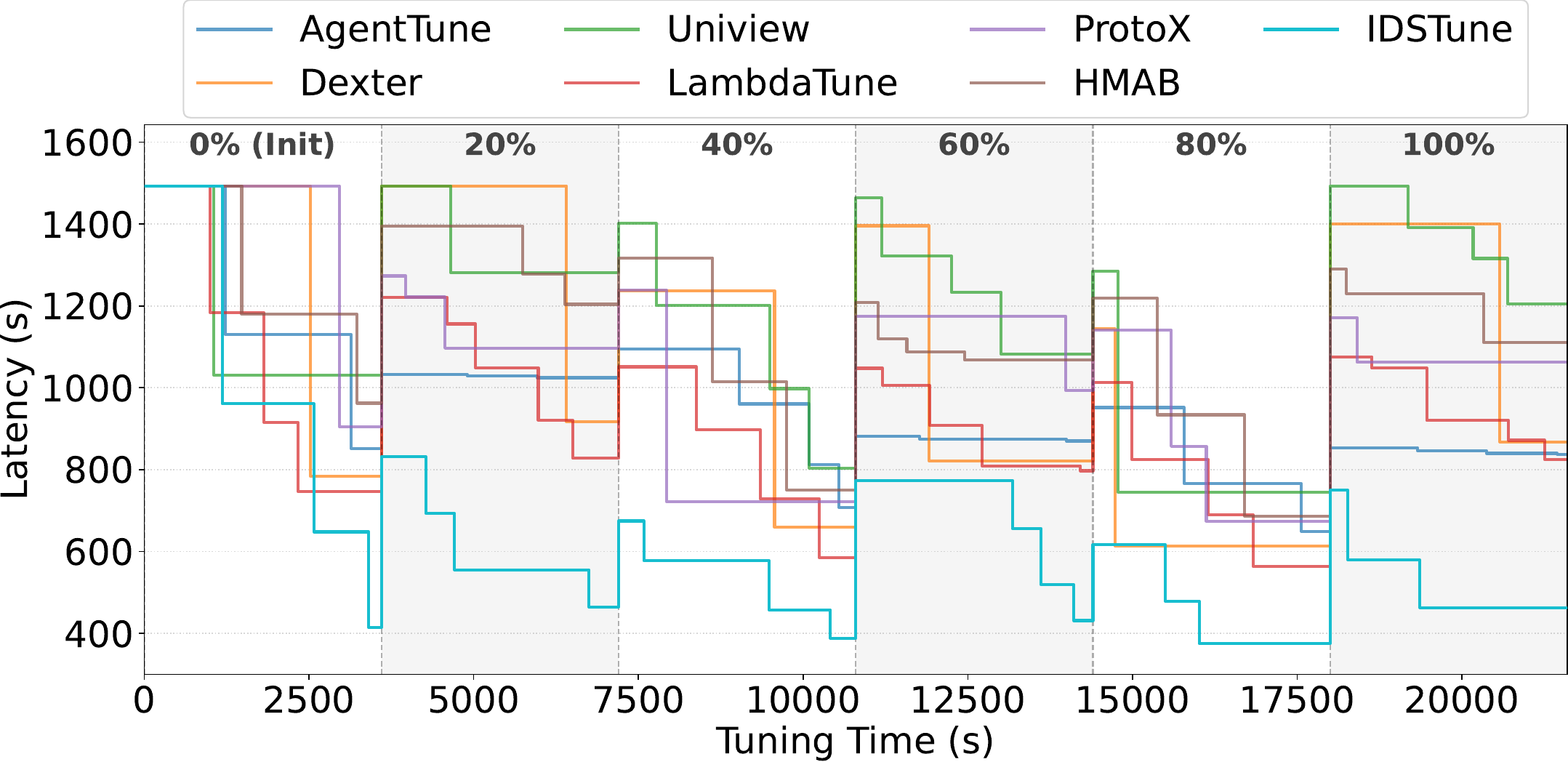}
    \vspace{-2em}
    \caption{Robustness analysis under progressive data drift.}
    \label{fig:drift}
  \end{minipage}
  \hfill 
  \begin{minipage}[t]{0.37\textwidth} 
    \centering
    \vspace{0pt}
    
    \captionof{table}{Scalability experiment.}
    \label{tab:db scaling}
    \vspace{-1em}
    \resizebox{\textwidth}{!}{
      \begin{tabular}{ccccc}
    \toprule
      Database Scale & \multicolumn{2}{c}{Latency (s)} & Runtime & Token  \\
      \cline{2-3}
      (GB) & Default & Optimized & Overhead (s) & Cost \\
    \midrule

    1  & 78.19 & 11.73  & 182.30 & 542.21K \\
    5  & 947.61 &  145.06 & 188.38 & 552.90K \\
    10  &  2800.15 & 311.64 & 174.08 & 537.10K  \\
    20  & 6013.04 & 635.95 & 180.47 & 563.54K  \\
  \bottomrule
\end{tabular}
    }
    
    \vspace{1em} 
    
    \captionof{table}{Ablation study on the workload compression.}
    \label{tab:ablation_compression}
    \vspace{-1em}
    \resizebox{\textwidth}{!}{%
\begin{tabular}{cccc}
    \toprule
      \makecell{Workload Compression \\ Method} & \makecell{Best-found \\ Performance (s) }& \makecell{Runtime\\ Overhead (s)} & \makecell{Token\\ Cost} \\
    \midrule
     Representation + Selection & \textbf{379.25} & 289.37 & 869.48K\\
     Representation Only & 667.48 & 1195.10 & 4611.94K \\
     No Workload & 799.53 & \textbf{27.59}  &  \textbf{12.94K} \\
  \bottomrule
\end{tabular}
    }
  \end{minipage}
  \vspace{-0.5em}
\end{figure*}

%% file: tables/ablation_multi-agent.tex
\begin{table}
\caption{Effectiveness of Multi-Agent Collaborative Tuning.}
\vspace{-1em}
\label{tab:ablation_framework}

\resizebox{0.46\textwidth}{!}{
\begin{tabular}{cccc}
    \toprule
    Method & \makecell{Best-found\\ Performance (s)} & \makecell{Runtime\\ Overhead (s)}  & \makecell{Token\\ Cost} \\
    \midrule
     Multi-Agent Collaborative (IDSTune) & \textbf{379.25} & 289.37 & 869.48K\\
     Multi-Agent Non-Collaborative & 493.02 & 183.04 & 582.56K\\
     Single-Agent & 617.37 & \textbf{133.12}  &  \textbf{508.66K} \\
  \bottomrule
\end{tabular}
}
\vspace{-1.2em}
\end{table}

%% file: tables/ablation_safety.tex
\begin{table}
\caption{\lyy{Ablation study on safety guardrails.}}
\vspace{-1em}
\label{tab:ablation_safety}

\resizebox{0.46\textwidth}{!}{
\begin{tabular}{ccccc}
    \toprule
    Method & \makecell{Best-found\\ Performance (s)} & \makecell{\lyy{\#False}\\ Negatives} & \makecell{\lyy{\#False}\\ Positives} &\makecell{Token\\ Cost} \\
    \midrule
     IDSTune & \textbf{379.25} & \textbf{0} & 2 & \textbf{869.48K}\\
      - w/o rule-based constraints & 399.35 & 1 & 3 & 871.64K\\
      - w/o LLM-based verification & 406.94 & 2 & \textbf{0} &  887.15K \\
      - w/o safety guardrails & 455.10 & 4 & 0 & 917.53K  \\
  \bottomrule
\end{tabular}
}
\vspace{-1em}
\end{table}

%% file: tables/ablation_search.tex
\begin{table}
\caption{Ablation study on the search mechanism.}
\vspace{-1em}
\label{tab:ablation_search}

\resizebox{0.46\textwidth}{!}{
\begin{tabular}{cccc}
    \toprule
      Operation Mode & \makecell{Best-found\\ Performance (s)} & \makecell{Runtime\\ Overhead (s)} & \makecell{Token\\ Cost} \\
    \midrule
    Forced-On  & \textbf{355.34} & 372.56 & 1083.22K\\
    Auto  & 379.25  & 289.37 & 869.48K \\
    Forced-Off  & 413.26 & \textbf{142.79} &  \textbf{635.54K} \\
  \bottomrule
\end{tabular}
}
\vspace{-0.5em}
\end{table}

%% file: appendix.tex
\section{In-Depth Analysis}
\renewcommand{\thefigure}{\thesection.\arabic{figure}}
\renewcommand{\thetable}{\thesection.\arabic{table}}

\setcounter{figure}{0}
\setcounter{table}{0}
\input{tables/case_study2}
We analyze the JOB benchmark on Postgres in more detail to better understand the performance gap observed in Figure~\ref{fig:main_result} of the main paper. In particular,  we examine a representative compositional baseline that sequentially combines UniView for materialized view selection and $\lambda$-Tune for knob and index tuning.

Table~\ref{tab:case_study2} reports the best configuration recommended by IDSTune and the baseline. For \textbf{knobs}, both approaches focus on improving data locality and enabling parallel query execution via setting 
\texttt{effective\_cache\_size}=12 GB, \texttt{max\_parallel\_workers}=8. IDSTune goes further by carefully tuning memory-related knobs (\texttt{shared\_buffers}=4 GB, \texttt{temp\_buffers}=128 MB) and worker settings (\texttt{max\_worker\_processes}=8) to match the hardware and workload characteristics. In addition, lowering \texttt{random\_page\_cost} to 2.5 steers the optimizer toward index-aware plans without excessively favoring index-driven nested-loop joins. Compared with the baseline, which aggressively reduces \texttt{random\_page\_cost} to 1.1 and increases \texttt{work\_mem} to 256 MB while leaving several other knobs at defaults, IDSTune adopts a more balanced cost model that is more robust for deep multi-way joins in the JOB workload.


The recommended \textbf{indexes} target frequent join and filter columns across tables such as \texttt{title}, \texttt{movie\_info}, \texttt{movie\_companies}, and \texttt{cast\_info}.
Composite indexes like:
\texttt{(movie\_id, info\_type\_id) in movie\_info}, 
\texttt{(movie\_id, company\_id) in movie\_companies}
directly optimize the most common join patterns observed in the workload.
In contrast, baseline methods typically construct a large number of single-column indexes to maximize coverage, which often results in higher maintenance overhead and reliance on bitmap operations. IDSTune instead prioritizes structure-aware composite indexes that align with join predicates, leading to more predictable execution plans.

IDSTune also produces five \textbf{materialized views} that precompute high-cost joins and frequently queried combinations, which frequently occur across analytical queries. For instance, \texttt{mv\_company\\\_keyword\_movie\_title} (View 1) pre-joins company, movie, and keyword relations, and \texttt{mv\_cast\_info\_name\_title} (View 3) resolves actor-role-movie relationships in advance. These MVs significantly shorten query execution paths and reduce repeated computation, yielding low-latency analytical responses even without additional hardware.
Compared with baseline materialized views that are often tied to specific constants or query predicates, IDSTune abstracts common join backbones of the JOB workload, enabling reuse across multiple queries rather than benefiting only a small subset.

Unlike traditional tuning methods that optimize knobs, indexes, or views independently, IDSTune explicitly considers the interactions among these configuration dimensions. For example, increasing \texttt{shared\_buffers} and \texttt{effective\_cache\_size} increases the utility of the newly added indexes, because more index blocks can now reside in memory.
In addition, enabling parallel workers provides the most benefit when heavy joins are shifted to pre-joined materialized views, reducing intermediate data exchange overhead. This coordinated reasoning enables IDSTune to achieve a globally balanced configuration, avoiding the local optima often observed in phase-wise baselines. 
In contrast, the baseline materializes only a subset of joins while aggressively biasing the optimizer toward index-driven plans, which can lead to suboptimal execution strategies for queries dominated by deep multi-way joins.

\begin{figure}
\vspace{-0.5em}
    \centering
    \includegraphics[width=0.48\textwidth]{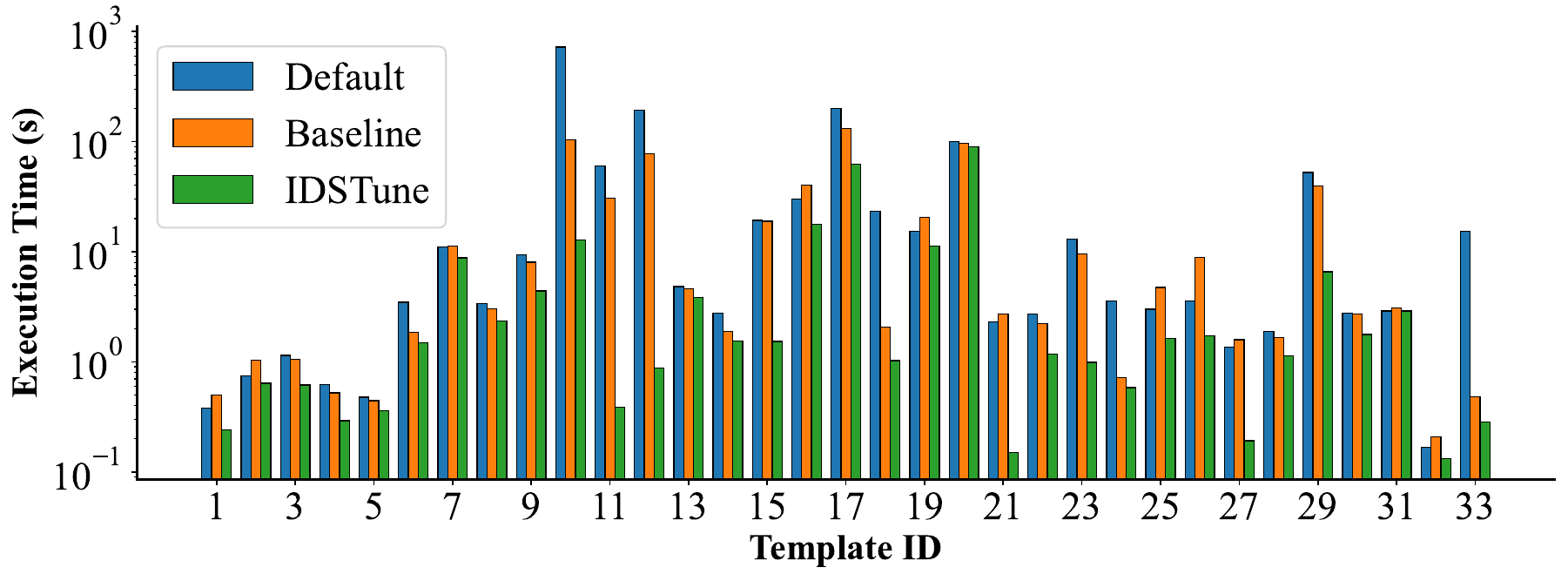}
    \caption{\lyy{Query execution times (JOB, Postgres).}} 
    \label{fig:JOB_compare}
    \vspace{-1em}
\end{figure}

\lyy{Finally, we compare per-query execution times across the default setting,  baseline method, and IDSTune for JOB. Figure~\ref{fig:JOB_compare} reports corresponding results. It turns out that the performance gain via the configuration proposed by IDSTune translate to gains or at least equal performance, compared to the default setting and baseline method, for each single query.}

\section{\lyy{Case Study: Multi-Agent Interaction in a Single Iteration}}
\lyy{We present a case study to illustrate how the proposed multi-agent framework performs coordinated tuning in a single iteration under the JOB workload. }

\vpara{Knob Specialist.} \lyy{The Knob Specialist agent generates an initial configuration focusing on memory allocation and cost modeling. For instance, it sets \texttt{shared\_buffers} to 2 GB and \texttt{effective\_cache\\\_size} to 12 GB to improve caching, while increasing \texttt{work\_mem} to 64 MB to accelerate hash joins. It also adjusts cost knobs (e.g., \texttt{random\_page\_cost}=2.0) to better reflect storage characteristics, thereby guiding the optimizer toward more sequential access patterns.}

\vpara{Index Specialist.} \lyy{Based on observed access patterns, the Index Specialist agent proposes indexes on frequently joined columns. In particular, it reinforces indexes on high-cardinality foreign keys such as movie\_id and person\_id in large tables (e.g., cast\_info, movie\_info), which exhibit substantial scan counts. These indexes aim to reduce join costs and improve access locality.}

\vpara{View Specialist.} \lyy{To further reduce repeated computation, the View Specialist agent introduces several materialized views that precompute common aggregation and join patterns. For example, it constructs views that aggregate keyword counts per movie and summarize company-level statistics, effectively caching intermediate results shared across queries.}

\vpara{Supervisor.} \lyy{The Supervisor agent then evaluates the combined recommendations holistically. In this iteration, the proposal is rejected due to suboptimal configuration settings and potential redundancies. Specifically, the reviewer identifies that (i) memory-related knobs (e.g., \texttt{shared\_buffers}, \texttt{work\_mem}) are too conservative for the workload, (ii) index recommendations lack composite coverage, and (iii) some materialized views may overlap with index benefits. The feedback is subsequently propagated back to individual agents, enabling targeted refinements in the next iteration.}

%% file: tables/case_study2.tex
\begin{table*}
\caption{Configuration Comparison between IDSTune and Baseline for JOB (Postgres).}
\vspace{-1em}
\label{tab:case_study2}

\resizebox{0.8\textwidth}{!}{
\begin{tabular}{|c|c|c|}
    \hline
      \textbf{Knob} & \textbf{\makecell{Recommended Value\\IDSTune}} & \ \textbf{\makecell{Recommended Value\\Baseline}} \\ \hline
      \texttt{\textbf{effective\_cache\_size}} & \textbf{12 GB} & \textbf{12 GB}\\ \hline
      \texttt{maintenance\_work\_mem} & 2 GB & 2 GB\\ \hline
      \texttt{\textbf{max\_parallel\_workers\_per\_gather}} & \textbf{4} & \textbf{4}\\ \hline
      \texttt{max\_parallel\_workers} & 8 & 8 \\ \hline
      \texttt{\textbf{work\_mem}} & \textbf{128 MB} & \textbf{256 MB}\\ \hline
      \texttt{\textbf{random\_page\_cost}} & \textbf{2.5} & \textbf{1.1} \\ \hline
        \texttt{seq\_page\_cost} & / & 1.0\\ \hline
      \texttt{checkpoint\_completion\_target} & / & 0.9\\ \hline
      \texttt{\textbf{max\_worker\_processes}} & \textbf{8} & \textbf{/} \\ \hline
      \texttt{max\_parallel\_maintenance\_workers} & 4 & / \\ \hline
      \texttt{\textbf{shared\_buffers}}    & \textbf{4 GB} & \textbf{/} \\ \hline
      \texttt{\textbf{temp\_buffers}} & \textbf{128 MB} & \textbf{/}\\ \hline
\end{tabular}
}

\vspace{1em}

\resizebox{0.92\textwidth}{!}{
\begin{tabular}{|c|c|c|}
    \hline
      \textbf{Table} & \textbf{\makecell{Recommended Indexes\\IDSTune}} & \textbf{\makecell{Recommended Indexes\\Baseline}} \\ \hline
      title    & production\_year, kind\_id, title & id\\ \hline
      movie\_info & (movie\_id, info\_type\_id), info & movie\_id, info\_type\_id\\ \hline
      movie\_info\_idx & (movie\_id, info\_type\_id), info & movie\_id, info\_type\_id\\ \hline
      movie\_companies & (movie\_id, company\_id), note & company\_id, company\_type\_id\\ \hline
      movie\_keyword & (movie\_id, keyword\_id) & movie\_id, keyword\_id\\ \hline
      cast\_info & (movie\_id, person\_id), person\_role\_id & movie\_id, person\_id\\ \hline
      company\_name & country\_code, name & / \\ \hline
      person\_info & (person\_id, info\_type\_id) & / \\ \hline
      movie\_link & (movie\_id, linkred\_movie\_id) & movie\_id, linked\_movie\_id\\ \hline
\end{tabular}
}

\vspace{1em}

\resizebox{0.92\textwidth}{!}{
\begin{tabular}{|c|c|}
    \hline
      \textbf{Materialized View-IDSTune} & \textbf{Materialized View-Baseline}  \\ \hline
      \makecell{SELECT cn.id AS company\_id, cn.country\_code, k.id AS keyword\_id,\\
      k.keyword, t.id AS movie\_id, t.title FROM company\_name cn JOIN \\movie\_companies mc ON cn.id = mc.company\_id JOIN title t ON \\mc.movie\_id = t.id JOIN movie\_keyword mk ON t.id = mk.movie\_id \\JOIN keyword k ON mk.keyword\_id = k.id;} & \makecell{SELECT keyword.id AS id, keyword.keyword AS \\keyword,
      movie\_keyword.movie\_id AS movie\_id \\from movie\_keyword,keyword \\where
      (movie\_keyword.keyword\_id = keyword.id) \\And (keyword.keyword = 'character-name-in-title')}\\ \hline
      \makecell{SELECT t.id AS movie\_id, t.title, t.production\_year, \\mi\_idx.info\_type\_id, mi\_idx.info FROM title t JOIN \\movie\_info\_idx mi\_idx ON t.id = mi\_idx.movie\_id;} & \makecell{SELECT info\_type.id AS id,\\ info\_type.info AS info,\\ movie\_info\_idx.movie\_id AS movie\_id}\\ \hline
      \makecell{SELECT ci.movie\_id, ci.person\_id, ci.role\_id, ci.note, n.name, \\n.gender, t.title, t.production\_year FROM cast\_info ci JOIN name \\n ON ci.person\_id = n.id JOIN title t ON ci.movie\_id = t.id} & \makecell{SELECT keyword.id AS id, keyword.keyword AS keyword,\\ movie\_keyword.movie\_id AS movie\_id, title.episode\_nr AS\\ episode\_nr,title.title AS title from title,movie\_keyword,\\keyword where (title.id = movie\_keyword.movie\_id) And\\ (movie\_keyword.keyword\_id = keyword.id) \\And (keyword.keyword = 'character-name-in-title')}\\ \hline
      \makecell{SELECT t.id AS movie\_id, t.title, t.production\_year, \\mi.info\_type\_id, mi.info FROM title t JOIN movie\_info mi ON t.id \\= mi.movie\_id;} & \makecell{SELECT keyword.id AS id, keyword.keyword AS keyword, \\movie\_companies.company\_id AS company\_id, \\movie\_companies.movie\_id AS movie\_id, title.title AS title from title,\\movie\_keyword,movie\_companies,keyword where (title.id = movie\_\\keyword.movie\_id) And (movie\_companies.movie\_id = title.id) \\And (movie\_keyword.keyword\_id = keyword.id) \\And (keyword.keyword = 'character-name-in-title')}\\ \hline
     \makecell{SELECT cc.movie\_id, cc.subject\_id, cc.status\_id, t.title, \\t.production\_year FROM complete\_cast cc JOIN title t ON \\cc.movie\_id = t.id;} & \makecell{SELECT keyword.id AS id, keyword.keyword AS keyword,\\ movie\_keyword.movie\_id
     AS movie\_id from movie\_keyword,\\keyword where (movie\_keyword.keyword\_id = keyword.id) \\And (keyword.keyword like '\%sequel\%') from info\_type,movie\_\\info\_idx; where (movie\_info\_idx.info\_type\_id = info\_type.id) \\And (info\_type.info = 'bottom 10 rank')} \\ \hline
\end{tabular}
}
\vspace{-1em}
\end{table*}